\pdfoutput=1
\documentclass[aps,prd,%
tightenlines,%
eqsecnum,%
floats,% 
notitlepage,%
nofootinbib,%
nobibnotes,%
longbibliography,% article titles in bibliography
amsfonts,amssymb,amsmath%
]{revtex4-1}

\usepackage{ifthen}
\makeatletter%
\usepackage{xstring}
\IfSubStr{\@classoptionslist}{preprint}%
{}%
{}%
\makeatother%

\newcommand{\arxiv}{1}

\usepackage{graphicx}
\usepackage[caption=false]{subfig}             % For side-by-side figures
\graphicspath{{Figures/}}

\usepackage{hyperref}

\usepackage{enumerate} % advanced enumerate environment
\def\p{\partial}
\usepackage[mathscr]{eucal}
\usepackage[latin1]{inputenc}
\newcommand{\ket}[1]{\mbox{$| {#1} \rangle$}}
\newcommand{\bracket}[2]{\mbox{$\langle {#1} \!\mid\! {#2} \rangle$}}

\newcommand{\melt}[3]{\mbox{$\langle {#1} | {#2} | {#3} \rangle$}}

\newcommand{\brasub}[2]{\ensuremath{{}_{#2}\kern-1pt\langle {#1} |}}
\newcommand{\ketsub}[2]{\ensuremath{| {#1} \rangle_{\kern-1pt #2}}}
\newcommand{\bracketsub}[4]{\ensuremath{{}_{#3}\kern-1pt\langle {#1} \!\mid\! {#2} \rangle_{\kern-1pt #4}}}
\newcommand{\ketbrasub}[4]{\ensuremath{| {#1} \rangle_{\kern-1pt #3}{}_{#4}\kern-1pt\langle {#2} |}}
\newcommand{\meltsub}[5]{\ensuremath{{}_{#4}\kern-1pt\langle {#1} \!\mid\! {#2} \!\mid\! {#3} \rangle_{\kern-1pt #5}}}

\newcommand{\rbrasub}[2]{\ensuremath{{}_{#2}\kern-1pt( {#1} |}}
\newcommand{\rketsub}[2]{\ensuremath{| {#1} )_{\kern-1pt #2}}}
\newcommand{\rbracketsub}[4]{\ensuremath{{}_{#3}\kern-1pt( {#1} \!\mid\! {#2} )_{\kern-1pt #4}}}
\newcommand{\rketbrasub}[4]{\ensuremath{| {#1} )_{\kern-1pt #3}{}_{#4}\kern-1pt( {#2} |}}
\newcommand{\rmeltsub}[5]{\ensuremath{{}_{#4}\kern-1pt( {#1} \!\mid\! {#2} \!\mid\! {#3} )_{\kern-1pt #5}}}

\def\d{\textrm{d}}

\def\lp{{l}_{p}}
\def\undertilde#1{\mathord{\vtop{\ialign{##\crcr
$\hfil\displaystyle{#1}\hfil$\crcr\noalign{\kern1.5pt\nointerlineskip}
$\hfil\tilde{}\hfil$\crcr\noalign{\kern1.5pt}}}}}

\begin{document}
\title{Cosmological dynamics in spin-foam loop quantum cosmology:
challenges and prospects}

\author{David A. Craig$^\star$ and Parampreet Singh$^\dagger$}
\affiliation{$^\star$ Department of  Physics, Le Moyne College\\
Syracuse, New York 13214, USA \\ $^\dagger$ Department of Physics and
Astronomy, Louisiana State University\\ Baton Rouge, Louisiana 70810, USA \\}

\begin{abstract}
We explore the structure of the spin foam-like vertex expansion in loop
quantum cosmology and discuss properties of the corresponding amplitudes, with
the aim of elucidating some of the expansion's useful properties and features.
We find that the expansion is best suited for consideration of conceptual
questions and for investigating short-time, highly quantum behavior.  In order
to study dynamics at cosmological scales, the expansion must be carried to
very high order, limiting its direct utility as a calculational tool for such
questions.  Conversely, it is unclear that the expansion can be truncated at 
finite order in a controlled manner.
% Ironically, the expansion turns out not to be a useful tool to investigate
% cosmological dynamics in the theory at any finite order.
\end{abstract}

% \pacs{98.80.Qc,04.60.Pp,04.60.Ds,04.60.Gw,04.60.Kz}  
% 03.65.Ca  Quantum Formalism 
% 03.65.Fd  Algebraic Methods
% 03.65.Ta  Foundations of quantum mechanics; measurement theory
% 03.65.Ud  Entanglement and quantum nonlocality
% 03.65.Yz  Decoherence
% 04.60.Ds  Canonical quantization
% 04.60.Gw  Covariant and sum-over-histories quantization
% 04.60.Kz  Minisuperspace models 
% 04.60.-m  Quantum gravity 
% 04.60.Nc  Lattice and discrete methods
% 04.60.Pp  Loop quantum gravity, quantum geometry
% 98.80.Qc  Quantum cosmology
% 
% \keywords{Quantum Cosmology, Minisuperspace, 
% Decoherence, Consistent Histories,
% % Generalized Quantum Theory, Alternative Formulations of Quantum Mechanics, 
% Loop Quantum Cosmology}

\maketitle

\section{Introduction}
\label{sec:intro}
Efforts to quantize Einstein's general relativity have fallen into two broad
classes, canonical approaches rooted in Dirac's quantization of constrained
systems, and covariant approaches based on the sum-over-histories formulation
of quantum theory.  Each of these approaches has respective strengths and
challenges.  Understanding the link between these distinct quantizations of
gravity and mapping the physical predictions between the canonical and the
covariant approaches is an important open problem.  Within the framework of 
the loop quantization of gravity, both paths to quantization have been 
pursued.
In the canonical loop quantum gravity approach, one aims to obtain a physical
Hilbert space with an inner product on physical states by implementing spatial
diffeomorphism and Hamiltonian constraints at the quantum level.  On the other
hand, in the covariant ``spin-foam'' formulation, the physical inner product
and the resulting physics are tied to summing over spin foam amplitudes
associated with a suitable discretization of the spacetime manifold.  Both
these approaches capture elements of a discrete quantum geometry, yielding
rich physical predictions.%
\footnote{For recent textbook-level introductions to these approaches, with 
many references to earlier literature, see \cite{CLLM,RovVid,kiefer12}.
} %
However, the precise relation between the canonical and covariant quantization
approaches in loop quantum gravity (LQG) is not yet clearly established.  The
primary reason lies in the mathematical complexities inherent in the
quantization of gravitational spacetimes, an as-yet incomplete task in both
approaches.  This naturally leads to the following questions: Can this
relationship be understood for simpler, yet still non-trivial spacetimes,
which \emph{can} be successfully quantized?  If so, how do we understand the
qualitative aspects of physics established in one approach in the framework of
the other approach?

To gain insight on these questions, cosmological spacetimes provide a useful
setting.  In the last decade, techniques of loop quantum gravity have been
applied to the successful quantization of various homogeneous cosmological
spacetimes, and the physical Hilbert space is known rigourously in loop
quantum cosmology (LQC) \cite{as-status,*liv-rev,*kinjal-rev,
*corichi-rev,*agullo-rev}.  A key prediction of LQC is the existence of a
``bounce'' when spacetime curvature becomes Planckian \cite{aps1,*aps2,*aps3}.
The existence of a bounce away from the curvature singularity has been
established in numerical simulations in a variety of models, even for highly
quantum states (for reviews see Refs.\ \cite{ps12,*khanna-review} and
\cite{numlsu-2,*numlsu-3}.)  Non-perturbative quantum gravitational effects
studied in a range of isotropic and anisotropic spacetimes point towards
resolution of all strong curvature singularities as a generic feature of LQC
\cite{ps9,*psvt, *ps11,*ps15, *ks-strong, *gowdy-sing}.  Further, for a
particular choice of lapse in the homogeneous and isotropic
Friedmann-Lema\^{i}tre-Roberston-Walker (FLRW) spacetime sourced with a
massless scalar field, the model can be solved exactly, establishing a robust
picture of the bounce for all the states in the physical Hilbert space
\cite{slqc}.  Using this solvable model of LQC (sLQC), self-consistent quantum
probabilities for the bounce can be calculated and the resolution of the
singularity demonstrated rigorously within the consistent histories
formulation of quantum theory \cite{consistent-lqc,*consistent,*craig-review}.

Interestingly, the solvable model can be used to give loop quantum cosmology a
spin-foam-like sum-over-histories formulation \cite{sflqc1,*sflqc2}.  The
theory may even be written as a standard path integral \cite{ach10b},
starting from the Hamiltonian in sLQC and obtaining a Lagrangian in the phase
space variables.%
\footnote{Note that the starting point of this spin-foam-like formulation of
loop quantum cosmology is not a covariant action in four dimensions, but
rather turns out be an infinite series in curvature invariants \cite{gops}.
} % 
This work provides a concrete platform from which to explore the physics of
canonical loop quantum cosmology in the covariant spin-foam language.  We will
refer to this formulation as spin-foam loop quantum cosmology, not to be
confused with spin-foam cosmology \cite{sf-cosmo1, sf-cosmo2, sf-sloan}, which
aims to investigate cosmological issues in the fully covariant spin-foam
formulation, albeit from a very different starting point and including certain
types of inhomogeneities.

Solvable LQC can be deparameterized, with the scalar field serving as a
physical matter ``clock''.  The quantum theory of sLQC can therefore be
analyzed, equivalently, using either a ``relativistic'' (Klein-Gordon) or
``non-relativistic'' (Schr\"{o}dinger) representation.  The latter allows a
direct connection with conventional Schr\"{o}dinger quantum mechanics and its
sum-over-histories formulation.  Nonetheless, a key difference from the
conventional path integral approach is that in sLQC one is dealing with
polymer quantization rather than standard Fock quantization.  The path
integral for LQC resembles the vertex expansion of spin-foam models obtained
by summing over suitably chosen dual triangulations which capture the
discretization of spacetime.  In work so far on the path integral formulation
of LQC \cite{sflqc1, sflqc2, twopoint, sflqc-b1, sflqc-b1m} the emphasis has
been on establishing the formal structure linking the two approaches.  This
has helped clarify certain technical issues in the spin-foam paradigm using
results from the canonical picture \cite{sflqc1,sflqc2}.  On the other hand,
using the covariant picture, insights into the choice of regulator and a local
vertex expansion have been obtained \cite{sflqc-b1m}.  Though these
developments have been useful, it has not been clear how effectively the
spin-foam like vertex expansion in LQC can be employed to shed light in a
practical way on cosmological dynamics in the spin-foam motivated language.
If it can be used, then one would like to understand features of the physical
evolution in the covariant picture.

In this manuscript, within the framework of spin-foam LQC we explore some
features of the vertex expansion of the theory's ``transition amplitudes''
(equivalently, inner product).  The vertex expansion is composed of a sum over
amplitudes for spacetime histories which undergo $M$ discrete volume
transitions.  The $M$th term in the vertex expansion in the spin-foam context
refers to a dual triangulation of the spacetime manifold with $M$ vertices.
We study the way amplitudes for different terms in the vertex expansion scale
with volume, and the way this scaling is affected by the degeneracies in the
volume transitions.  The behavior for the small $M$ cases is found by explicit
computation.  We argue that in order to capture cosmologically relevant
physical evolution, large orders $M$ in the expansion are required.  We show
in particular that low orders in the vertex expansion can at best capture the
dynamics of only short intervals $\Delta\phi$ in the matter field $\phi$ and
small changes in cosmological volume.  In that regard the vertex expansion in
LQC is much like an expansion of a standard quantum propagator in powers of
$Et/\hbar$.  Truncating the expansion at low order yields an approximation
that is useful to study physical evolution only for short time intervals.  A
different expansion in LQC will therefore be necessary in order to study
cosmological dynamics.

Moreover, we also find that the while the quantum amplitudes do satisfy the
quantum constraint, the vertex expansion of those amplitudes does not satisfy
the constraint in a controlled way if it is truncated at any finite order,
once again calling into question the utility of the vertex expansion in
studying the theory's cosmological dynamics.

The manuscript is organized as follows.  In the next section, we summarize the
main results from solvable LQC and discuss the quantum Hamiltonian constraint
and its different representations.  In this manuscript we focus on the
``relativistic'' representation, which results in a representation for the
quantum transition amplitudes that is directly analogous to the Hadamard
propagator of standard quantum field thoery, as discussed in Sec.\ III. Some
useful properties and a reformulation of the vertex expansion are discussed in
Sec.  IV.  In Sec.\ V we discuss various results.  We conclude with a
summary in Sec.\ VI.

% This physical inner product, which can be thought analog of the transition
% amplitude in the conventional path integral formulation, is

% It is expected that for a quantization of a given gravitational spacetime,
% both approaches result in a similar physics and provide complementary
% insights on the nature of quantum spacetime.  It is also hoped that one
% should be able to derive results of one approach using another through a
% well defined procedure.  In order to practically answer these questions and
% gain insights on the underlying issues, cosmological spacetimes provide a
% very useful arena.

\section{Solvable loop quantum cosmology}
\label{sec:sLQC}

% By and large, the conventions employed will be as in \cite{dac13a,CS13a}.
% Note we take everywhere $c=1$ so $l_p = \sqrt{G\hbar/c^3} =\sqrt{G\hbar}$.

The quantization described in our analysis is based on the spacetime metric
for the spatially flat, homogeneous isotropic spacetime, given by
\begin{equation} 
\d s^2 = -N^2 \d t^2 + q_{ab} \d x^a \d x^b = -N^2 \d t^2 + a^2(t) d {\bf x}^2
~,
\end{equation} 
where $N(t)$ is the lapse function.  The spatial topology is taken to be
$\mathbb{R}^3$ and a fiducial spatial cell ${\cal V}$ is fixed in order to
define a symplectic structure after integration over the spatial volume of
${\cal V}$.  We take the matter to be a massless, minimally coupled scalar
field $\phi$.  The classical cosmological dynamics for this spacetime and
matter content yields expanding and contracting solutions which are disjoint
and singular for any given value of the scalar field momentum $p_\phi$.  The
loop quantization for this metric for lapse $N=1$ was first rigorously
performed in Refs.  \cite{aps1, *aps2, *aps3}.  The quantization results in
resolution of the cosmological singularity at the level of the physical
Hilbert space.  The big bang and big crunch singularities in the expanding and
contracting branches are avoided, and replaced by a non-singular ``bounce'' of
the universe at the Planck scale and effective unitary evolution if the scalar
field $\phi$ is taken as a physical clock.  Significant control over the
physical Hilbert space can be achieved by choosing the lapse $N = a^3$, in
which case the model becomes exactly solvable (``sLQC'') \cite{slqc}.  In the
following we will concern ourselves with this particular choice, and summarize
some of the basic features of the construction.

At the classical level, after imposition of the symmetries of this spacetime,
the symmetry reduced gravitational phase space variables are the connection
$c$ and the triad $p$, and the classical Hamiltonian constraint can be written
as
\begin{equation}  
{C} = p_\phi^2 - 3 \pi G v^2 b^2 \approx 0 ~.
\end{equation} 
Here $v$ is related to the physical volume $V = |p|^{3/2}$ of the fiducial
cell ${\cal V}$ as $v = V/2 \pi G$, and $b = c/|p|^{1/2}$.  The modulus sign
arises due to the two possible orientations of the triad.  The phase space
variables are $(v,b)$ and $(\phi, p_\phi)$, satisfying
\begin{equation} 
\{v, b \} = 2 \gamma, ~~~~\mathrm{and} ~~~~ \{\phi, p_\phi\} = 1 ~,
\end{equation} 
with $\gamma \approx 0.2375$ as the Barbero-Immirzi parameter.  Upon
quantization, the action of the volume operator is multiplicative on states
$\Psi(\nu)$,
\begin{equation} 
\hat V \, \Psi(\nu) = 2 \pi \gamma \lp^2 |\nu| \Psi(\nu) ~,
\end{equation} 
where $\nu = v/\gamma \hbar$.  Unlike $V$, there is no corresponding operator
$\hat b$ in the loop quantization.  Rather, its action is captured via
holonomies of the connection, through the translation operator acting on the
volume eigenkets $|\nu \rangle$:
\begin{equation} 
\widehat{\exp(i \lambda b)} |\nu\rangle = |\nu - 2 \lambda\rangle ~,
\end{equation} 
where $\lambda$ captures the minimum non-zero area in quantum geometry given
by $\lambda^2 = 4\pi\sqrt{3}\gamma \lp^2$.  This action of the holonomy
operators is responsible for the discrete quantum evolution discussed
below.

Using the action of $\hat p_\phi=-i\hbar\partial_{\phi}$ on states
$\Psi(\nu,\phi)$, the quantum Hamiltonian constraint can be written in a
Klein-Gordon form,
\begin{equation} \label{eqC}
\hat{C} \Psi(\nu,\phi) = - (\partial_{\phi}^2 + \Theta(\nu)) \Psi(\nu,\phi) ~.
\end{equation} 
Here $\Theta$ is a positive definite and essentially self-adjoint operator,
with continuous eigenvalues $\omega_k = \kappa |k|$, where $\kappa = \sqrt{12
\pi G}$:
% \begin{align*}
% \Theta\,\Psi(\nu,\phi) & =  \\
%    -\frac{3\pi G}{4\lambda^2}& \left\{
% \sqrt{|\nu(\nu+4\lambda)|}|\nu+2\lambda|\Psi(\nu+4\lambda,\phi) - 2\nu^2\Psi(\nu,\phi)
% + \sqrt{|\nu(\nu-4\lambda)|}|\nu-2\lambda|\Psi(\nu-4\lambda,\phi)
% \right\}
% \end{align*}
\begin{equation} 
\Theta\ket{k} = \omega_k^2\ket{k} ~.
\end{equation} 
The symmetric eigenfunctions of $\Theta$ are $e^{(s)}_{k}(\nu)\equiv
\langle\nu|k\rangle$.  They are orthogonal and satisy the completeness
relations
\begin{eqnarray}
\sum_{\nu=4\lambda n} e^{(s)}_{k}(\nu)^* e^{(s)}_{k'}(\nu) & = & \delta^{(s)}(k,k') ~,
\label{eq:eskcompletek}\\
\int_{-\infty}^{+\infty}dk\, e^{(s)}_{k}(\nu) e^{(s)}_{k}(\nu')^* & = &
\delta^{(s)}_{\nu,\nu'}  ~.
\label{eq:eskcompletev}
\end{eqnarray}
(See \cite{craig-eigen} for further details.)

On physical states $\Psi(\nu,\phi)$ the action of $\Theta$ is 
\begin{equation} \label{eq:theta}
\Theta\,\Psi(\nu,\phi) =
   -\frac{3\pi G}{4\lambda^2} \left\{
\sqrt{|\nu(\nu+4\lambda)|}|\nu+2\lambda|\Psi(\nu+4\lambda,\phi) - 2\nu^2\Psi(\nu,\phi)
+ \sqrt{|\nu(\nu-4\lambda)|}|\nu-2\lambda|\Psi(\nu-4\lambda,\phi)
\right\} ~.
\end{equation} 
The corresponding matrix elements may be written
\begin{equation} \label{eq:thetavv}
\Theta_{\nu\nu'} = \left(\frac{\kappa}{4\lambda}\right)^2
  \sqrt{|\nu\cdot\nu'|}|\nu+\nu'|
  \left\{\delta_{\nu,\nu'}-\frac{1}{2}[\delta_{\nu+4\lambda,\nu'}+\delta_{\nu-4\lambda,\nu'}]\right\} ~.
\end{equation}
In this simple model, $\Theta$ is a $\phi$-independent 
% time-independent
spatial Laplacian operator whose action links the physical wavefunction with
uniform discreteness in volume.  The physical Hilbert space can be decomposed
into disjoint sectors of positive and negative frequency symmetric solutions
to the quantum constraint which satisfy
\begin{equation} 
\mp i \p_\phi \Psi(\nu,\phi) = \sqrt{\Theta} \Psi(\nu,\phi) ~. \label{posfreq}
\end{equation} 
A physical state has support only on a lattice $\epsilon \in [0,4)$ related to
$\nu$ via $\nu = 4 n \lambda + \epsilon$ where $n \in \mathbb{Z}$.  Each
lattice is left invariant by the unitary dynamical evolution in $\phi$.  This
results in a super-selection of the physical Hilbert space: ${\cal
H}_{\mathrm{phys}} = \oplus {\cal H}_\epsilon$.  In our analysis we will work
with the $\epsilon = 0$ lattice which includes the case of zero volume -- the
classical big bang singularity.  The classical singularity is resolved in the
quantum theory leading to a ``bounce'' at small volume
\cite{aps1,aps2,aps3,slqc}, where quantum geometry leads to an effective
repulsive force.  The bounce can also be understood via properties of the
eigenfunctions of $\Theta$ numerically \cite{aps3} as well as analytically
\cite{craig-eigen}.  The eigenfunctions are found to decay exponentially near
the classical singularity unless $k \lesssim |\nu|/2\lambda$.  The exponential
fall-off is determined by the value of the scalar field momentum which is a
direct measure of the bounce volume.

Eq.(\ref{posfreq}) is analogous to the Schr\"{o}dinger equation of ordinary
non-relativistic quantum mechanics.  In this representation, positive and
negative frequency states can be expanded in terms of the symmetric
eigenfunctions of the $\Theta$ operator as
\begin{equation} 
\Psi^{\pm}(\nu,\phi) = \int_{-\infty}^{\infty}dk\, \tilde{\Psi}(k)\, 
e^{(s)}_{k}(\nu)\, e^{\pm i\omega_k\phi} ,
\end{equation} 
where $\Psi(k)$ is the wave profile.  The physical states have finite norm
with respect to the Schr\"{o}dinger inner product computed at a fiducial (but
immaterial) $\phi = \phi_o$,
\begin{equation} 
\langle \Psi_1|\Psi_2\rangle = \sum_{\nu=4n} \Psi^*_1(\nu,\phi_o) \Psi_2( \nu,\phi_o) ~.
\end{equation} 
A unitarily equivalent representation is the ``relativistic'' one where the physical
states can be written as
\begin{equation} 
\undertilde{\Psi}^{\pm}(\nu,\phi) = \int_{-\infty}^{\infty}\frac{dk}{\sqrt{2\omega_k}}\, \tilde{\Psi}(k)\, % e^{i\omega_k\phi}\, e^{(s)}_{k}(\nu)
e^{(s)}_{k}(\nu)\, e^{\pm i\omega_k\phi} ~.
\end{equation} 
In this representation the inner product is the Klein-Gordon product
\begin{equation} 
\bracket{\undertilde{\Psi}^{\pm}}{\, \undertilde{\Phi}^{\pm}} =
\mp i\sum_{\nu=4n\lambda} 
\undertilde{\Psi}^{\pm}(\nu,\phi)^*\buildrel{\leftrightarrow}\over{\partial}_{\phi}\undertilde{\Phi}^{\pm}(\nu,\phi),
\end{equation} 
which for the current simple model (zero potential) turns out to be precisely
equal to the Schr\"{o}dinger inner product.  Here
$\buildrel{\leftrightarrow}\over{\partial}_{\phi} =
\buildrel{\rightarrow}\over{\partial}_{\phi} -
\buildrel{\leftarrow}\over{\partial}_{\phi}$.  Note that the
$\sqrt{2\omega_k}$ in the measure could alternatively be absorbed into a
renormalization of the eigenfunctions and corresponding completeness
relations.  To make the connection with the covariant description and full
quantum gravity, this ``relativistic'' representation is more natural.
Further, it is a useful representation to work with for more general models
which do not deparameterize.  In the following analysis, we will primarily
employ the relativistic representation.

\section{Propagators}
\label{sec:reps}
An object of primary interest in our discussion is the ``extraction
amplitude'' 
-- essentially, the inner prouct --  %
which in the relativistic representation turns out to be in essence the
Hadamard propagator of ordinary quantum field theory, and which can be
interpreted as a propagator in LQC as well.  On the other hand, in the
non-relativistic representation the extraction amplitude is the Newton-Wigner
function.  In the following, we summarize the construction behind these
amplitudes and obtain the composition laws which enable them to be viewed as
propagators.  Our discussion will be based on the earlier analyses of
Refs.~\cite{sflqc1,sflqc2,twopoint}.

Let us start with the Hadamard function, a two point function in the
relativistic representation, given by the physical inner product between
the eigenstates $\ket{\nu,\phi}$, which can be obtained using group averaging:
\begin{eqnarray} 
G_{\mathrm{H}}(\nu_f,\phi_f;\nu_i,\phi_i) &\equiv&
\bracket{\nu_f,\phi_f}{\nu_i,\phi_i}\nonumber\\
&=&
% \meltsub{\nu_f,\phi_f}{\delta(\hat{C})}{\nu_i,\phi_i}{\mathrm{kin}}{\mathrm{kin}} \\
% \rmelt{\nu_f,\phi_f}{\delta(\hat{C})}{\nu_i,\phi_i} \\
% &=&
\int_{-\infty}^{\infty}d\alpha\
\meltsub{\nu_f,\phi_f}{e^{i\alpha\hat{C}}}{\nu_i,\phi_i}{\mathrm{kin}}{\mathrm{kin}}, 
% \rmelt{\nu_f,\phi_f}{e^{i\alpha\hat{C}}}{\nu_i,\phi_i} \\
\end{eqnarray} 
where $\hat C$ is given by (\ref{eqC}), $\alpha$ is the group averaging
parameter, and $\ketsub{\nu,\phi}{\mathrm{kin}}$ denote the states in the
kinematical Hilbert space.  Since $\hat p_\phi$ commutes with the
$\phi$-independent $\Theta$, the Hadamard function can be written as a product
of two amplitudes,
\begin{equation} \label{gravamp}
A_{\mathrm{H}}(\Delta\phi;\alpha) =
\melt{\phi_f}{e^{i\alpha p_{\phi}^2}}{\phi_i}, ~~~\mathrm{and} ~~~
A_{\Theta}(\nu_f,\nu_i;\alpha) = \melt{\nu_f}{e^{-i\alpha\Theta}}{\nu_i} ~,
\end{equation} 
such that 
\begin{equation} 
G_{\mathrm{H}}(\nu_f,\phi_f;\nu_i,\phi_i) = \int_{-\infty}^{\infty}d\alpha\
A_{\mathrm{H}}(\Delta\phi;\alpha) A_{\Theta}(\nu_f,\nu_i;\alpha) ~.
\end{equation} 

Using $\bracket{\phi}{p_{\phi}} = \exp(ip_{\phi}\phi/\hbar)/\sqrt{2\pi}$ and
the resolution of the identity we get
\begin{equation} 
A_{\mathrm{H}}(\Delta\phi;\alpha)   = 
\int_{-\infty}^{\infty}dp_{\phi}\,
\melt{\phi_f}{e^{i\alpha p_{\phi}^2}}{p_{\phi}}\bracket{p_{\phi}}{\phi_i} = 
\int_{-\infty}^{\infty}\frac{dp_{\phi}}{2\pi}\,e^{i\alpha p_{\phi}^2}e^{ip_{\phi}\Delta\phi} ~.
\end{equation} 

Similarly, the gravitational part of the amplitude can be written as 
\begin{equation} 
A_{\Theta}(\nu_f,\nu_i;\alpha) = \int_{-\infty}^{\infty}dk\, \melt{\nu_f}{e^{-i\alpha\Theta}}{k}\bracket{k}{\nu_i} = 
 \int_{-\infty}^{\infty}dk\, e^{-i\alpha\omega_k^2}\, e^{(s)}_{k}(\nu_f) e^{(s)}_{k}(\nu_i)^* ,
\end{equation} 
where in the last step we have used the eigenvalue equation for $\Theta$. 

Using the above expressions for $A_{\mathrm{H}}$ and $A_\Theta$, we can
separate the Hadamard function into positive and negative frequency parts.  To
prove this, let us rearrange the integrals in $G_H$ as
\begin{equation} 
G_{\mathrm{H}}(\nu_f,\phi_f;\nu_i,\phi_i) = \int_{-\infty}^{\infty}dk\,  e^{(s)}_{k}(\nu_f) e^{(s)}_{k}(\nu_i)^*
\int_{-\infty}^{\infty}\frac{dp_{\phi}}{2\pi}\, e^{ip_{\phi}\Delta\phi}\,
\int_{-\infty}^{\infty}d\alpha\,  e^{-i\alpha(p_{\phi}^2-\omega_k^2)} ~.
\end{equation} 
Performing the integration over the group averaging parameter yields a sum of
Dirac delta functions which separate the terms with positive and negative
frequencies, resulting in
\begin{eqnarray}\label{eq:GH} 
G_{\mathrm{H}}(\nu_f,\phi_f;\nu_i,\phi_i) &=&
% \int_{-\infty}^{\infty}d\alpha\, \melt{\nu_f,\phi_f}{e^{i\alpha\hat{C}}}{\nu_i,\phi_i} \\
% &=& \int_{-\infty}^{\infty}d\alpha\,
% A_{\mathrm{H}}(\Delta\phi;\alpha) A_{\Theta}(\nu_f,\nu_i;\alpha) \\
% \int_{-\infty}^{\infty}dk\,  e^{(s)}_{k}(\nu_f) e^{(s)}_{k}(\nu_i)^*
% \int_{-\infty}^{\infty}\frac{dp_{\phi}}{2\pi}\, e^{ip_{\phi}\Delta\phi}\
% \frac{1}{2\omega_k} [\delta(p_{\phi}+\omega_k)+\delta(p_{\phi}-\omega_k)]\\
% &=&
\int_{-\infty}^{\infty}\frac{dk}{2\omega_k}\,
[e^{+i\omega_k\Delta\phi} + e^{-i\omega_k\Delta\phi}]\,
e^{(s)}_{k}(\nu_f) e^{(s)}_{k}(\nu_i)^*  \nonumber\\
&=&
\bracketsub{\nu_f,\phi_f}{\nu_i,\phi_i}{+}{+}
  + \bracketsub{\nu_f,\phi_f}{\nu_i,\phi_i}{-}{-} \nonumber\\
% &=&
% 2 \int_{-\infty}^{\infty}\frac{dk}{2\omega_k}\,
% \cos(\omega_k\Delta\phi) e^{(s)}_{k}(\nu_f) e^{(s)}_{k}(\nu_i)^*  \\
&=&
G^+_{\mathrm{H}}(\nu_f,\phi_f;\nu_i,\phi_i) + G^-_{\mathrm{H}}(\nu_f,\phi_f;\nu_i,\phi_i) ~.
\end{eqnarray} 
Here the ``Wightman functions'' $G^{\pm}_{\mathrm{H}}$ give the physical inner
products between the positive/negative frequency eigenstates.
% The
% third-to-last line is the key one for drawing conclusions about the behavior
% of $G_{\mathrm{H}}$ and/or $G^{\pm}_{\mathrm{H}}$.
Using the Klein-Gordon inner product, it is straightforward to show that they
satisfy the following composition law:
\begin{eqnarray} 
G^\pm_{\mathrm{H}}(\nu_f,\phi_f;\nu_i,\phi_i)  &=&
G^\pm_{\mathrm{H}}(\nu_f,\phi_f;\nu,\phi)  \circ 
  G^\pm_{\mathrm{H}}(\nu,\phi;\nu_i,\phi_i) \nonumber\\
&=&
\mp i \sum_{\nu}
G^\pm_{\mathrm{H}}(\nu_f,\phi_f;\nu,\phi)
\buildrel{\leftrightarrow}\over{\partial}_{\phi} G^\pm_{\mathrm{H}}(\nu,\phi;\nu_i,\phi_i) ~,
\end{eqnarray} 
where we have used the completeness relation for the eigenfunctions
$e^{(s)}_{k}(\nu)$.  (The first line defines the relativistic composition
operator $\circ$.)
% -- essentially the relativistic Klein-Gordon product.)  
This composition law allows us to view the Hadamard two point function as a
transition amplitude or propagator of the dynamics from $\phi_i$ to $\phi_f$.
The propagation action from the state $\Psi^{\pm}(\nu',\phi')$ to
$\Psi^{\pm}(\nu,\phi)$ is
\begin{equation} 
\Psi^{\pm}(\nu,\phi) = G^\pm(\nu,\phi;\nu',\phi') \circ \Psi^{\pm}(\nu',\phi') 
= \mp i \sum_{\nu'}
G^\pm_{\mathrm{H}}(\nu,\phi;\nu',\phi')
\buildrel{\leftrightarrow}\over{\partial}_{\phi'} \Psi^{\pm}(\nu',\phi') ~.
\end{equation} 

% Interestingly, $G_H^+$ and $G_H^-$ provide a causal propagator
% $G_{\mathrm{c}}$ and the Feynman propagator $G_{\mathrm{F}}$ as follows:
% \begin{equation} G_{\mathrm{c}}(\nu,\phi;\nu',\phi') = -i(
% G^+_{\mathrm{H}}(\nu,\phi;\nu',\phi') -
% G^-_{\mathrm{H}}(\nu,\phi;\nu',\phi')) \end{equation} and \begin{equation}
% G_{\mathrm{F}}(\nu,\phi;\nu',\phi') = -i(
% G^+_{\mathrm{H}}(\nu,\phi;\nu',\phi')\theta(\phi-\phi') +
% G^-_{\mathrm{H}}(\nu,\phi;\nu',\phi')\theta(\phi'-\phi)) .  \end{equation}

One can similarly define a Newton-Wigner two-point function 
\cite{sflqc2,twopoint},
\begin{equation} 
G_{\mathrm{NW}}(\nu_f,\phi_f;\nu_i,\phi_i) \equiv
\int_{-\infty}^{\infty}d\alpha\
\meltsub{\nu_f,\phi_f}{e^{i\alpha\hat{C}}\, 2|p_{\phi}|}{\nu_i,\phi_i}{\mathrm{kin}}{\mathrm{kin}} ~.
\end{equation} 
Using the observation that the gravitational part of the amplitude $A_\Theta$
is identical to the one in the Hadamard case, it is easily shown that
the positive frequency/negative frequency pieces of the Newton-Wigner
function are related to those of the Hadamard propagator by
\begin{equation} 
G^\pm_{\mathrm{NW}}(\nu_f,\phi_f;\nu_i,\phi_i) =
\mp 2i\, \partial_{\phi_f}G^\pm_{\mathrm{H}}(\nu_f,\phi_f;\nu_i,\phi_i).
\end{equation} 
Note that $G_{\mathrm{NW}}$ is given by essentially the same expression as for
the Hadamard propagator, but without the $1/2\omega_k$ in the measure.

It is straightforward to check that it satisfies the following
``non-relativistic''composition law:
\begin{eqnarray} 
G^\pm_{\mathrm{NW}}(\nu_f,\phi_f;\nu_i,\phi_i)  &=& \nonumber 
G^\pm_{\mathrm{NW}}(\nu_f,\phi_f;\nu,\phi)  \circ_{{\scriptscriptstyle \mathrm{NW}}} 
     G^\pm_{\mathrm{NW}}(\nu,\phi;\nu_i,\phi_i) \\
&=&
\sum_{\nu} G^\pm_{\mathrm{NW}}(\nu_f,\phi_f;\nu,\phi)\, G^\pm_{\mathrm{NW}}(\nu,\phi;\nu_i,\phi_i) ~.
\end{eqnarray} 
The Newton-Wigner propagator naturally propagates states in the 
Schr\"{o}dinger inner product,
\begin{equation} 
\Psi^{\pm}(\nu,\phi) = \sum_{\nu'}
G^\pm_{\mathrm{NW}}(\nu,\phi;\nu',\phi')\, \Psi^{\pm}(\nu',\phi').
\end{equation} 
The propagation relations for the Hadamard and Newton-Wigner functions show that
the propagation action is unchanged in both the ``non-relativistic'' and
``relativistic'' representations.

With these properties established, we are now equipped to employ the Hadamard
propagator to compute the vertex amplitude in spin-foam loop quantum
cosmology in the next section.

\section{The vertex expansion in spin foam loop quantum cosmology}
\label{sec:sflqc}
In order to compute the transition amplitude from a kinematical state
$|\nu_i,\phi_i\rangle$ to $|\nu_f,\phi_f\rangle$, we need to compute the
amplitude corresponding to the gravitational part given by
Eq.~(\ref{gravamp}).  To evaluate it, following \cite{sflqc1} we use the ideas
of the sum-over-histories formulation of quantum theory with $\Theta$
% $\alpha \Theta$
playing the role of the Hamiltonian.  The main idea behind this construction
is as follows.  The ``time'' interval $\alpha$
% $\Delta \phi = \phi_f - \phi_i$ 
is
divided into $N$ equal parts of length $\varepsilon$.  Each interval of time
is labelled by a volume element $\nu$ via insertions of resolutions of the
identity in volume.  This provides a time discrete history, $\Delta \phi_N$,
identified with $N-1$ volumes between $\nu_i$ and $\nu_f$.  The gravitational
part of the amplitude can then be written as
\begin{equation} 
A_\Theta(\nu_f,\nu_i;\alpha) = \sum_{\nu_{N-1}, ..., {\nu_1}} \langle\nu_N|e^{-i \varepsilon \Theta} |\nu_{N-1}\rangle ...  \langle\nu_1|e^{-i \varepsilon \Theta} |\nu_{0}\rangle =   
\sum_{\Delta \phi_N} U_{\nu_N \nu_{N-1}} ...  U_{\nu_1 \nu_{0}} ~.
% A_\Theta(\nu_f,\nu_i;\alpha) = \sum_{\nu_{N-1}, ..., {\nu_1}} \langle\nu_N|e^{-i \varepsilon \alpha \Theta} |\nu_{N-1}\rangle ...  \langle\nu_1|e^{-i \varepsilon \alpha \Theta} |\nu_{0}\rangle =   
% \sum_{\Delta \phi_N} U_{\nu_N \nu_{N-1}} ...  U_{\nu_1 \nu_{0}} ~.
% A_\Theta(\nu_f,\nu_i;\Delta \phi) = \sum_{\nu_{N-1}, ..., {\nu_1}} \langle\nu_N|e^{-i \varepsilon \alpha \Theta} |\nu_{N-1}\rangle ...  \langle\nu_1|e^{-i \varepsilon \alpha \Theta} |\nu_{0}\rangle =   
% \sum_{\Delta \phi_N} U_{\nu_N \nu_{N-1}} ...  U_{\nu_1 \nu_{0}} ~.
% \sum_{\nu_{N-1}, ..., {\nu_1}} U_{\nu_N \nu_{N-1}} ...  U_{\nu_1 \nu_{0}} ~.
\end{equation} 
where $\nu_0 = \nu_i$ and $\nu_N = \nu_f$.  In ordinary quantum mechanics, the
transition amplitude is then computed by taking $N \rightarrow \infty$ which
removes any dependence on 
$\varepsilon = \alpha/N$.  
% $\varepsilon = \Delta \phi/N$.  
Here this limit is
tricky, since each term in the above product yields an $\varepsilon$ term in
the first order.  
% This is because the lowest order identity term in the expansion of each of the 
% exponentials gives zero unless \nu_{i+1}=\nu_i, and hence is only non-zero 
% if all the volumes are equal, and in particular \nu_i=\u_f.$
The total product is thus proportional to $\varepsilon^N$, which vanishes in
the naive limit $N \rightarrow \infty$.  To take this limit, the above sum is
instead reorganized according to the number $M$ of discrete volume
\emph{transitions} to a distinct volume, regardless of ``when'' (i.e.\ at what
value of $\phi$) they occur.  The gravitational amplitude can then be written
as a sum over amplitudes for individual paths
$(\nu_{M},\nu_{M-1},\ldots,\nu_{1},\nu_{0})$ with $M$ transitions
\cite{sflqc1},
\begin{equation} \label{eq:vertexexptheta}
A_\Theta(\nu_f,\nu_i; \alpha) = \sum_{M=0}^N 
   \sum_{ \substack{\nu_{M-1},\ldots,\nu_1\\ \nu_m\neq \nu_{m+1}} }
A(\nu_{M},\nu_{M-1},\ldots,\nu_{1},\nu_{0};\alpha) ~.
% A_\Theta(\nu_f,\nu_i; \Delta \phi) = \sum_{M=0}^N 
%    \sum_{ \substack{\nu_{M-1},\ldots,\nu_1\\ \nu_m\neq \nu_{m+1}} }
% A(\nu_{M},\nu_{M-1},\ldots,\nu_{1},\nu_{0};\Delta \phi) ~.
\end{equation} 
This reorganization of the sum allows taking the limit $N \rightarrow \infty$,
which then results in the transition amplitude between $|\nu_i,\phi_i\rangle$
and $|\nu_f,\phi_f\rangle$ after integration over $\alpha$ and $p_{\phi}$.
Note that in this reorganization in terms of volume transitions, while by
construction no two \emph{consecutive} volumes will be the same, for any given
history $(\nu_{M},\nu_{M-1},\ldots,\nu_{1},\nu_{0})$ individual volumes may be
repeated.  Take $p$ to be the number of \emph{unique} volumes
$(w_{p-1},w_{p-2},\ldots,w_{1},w_{0})$ appearing in the path (so $p\leq M+1$),
where $w_0=\nu_0$.%
\footnote{Note this is a slight change from the notation of
\cite{sflqc1,sflqc2,twopoint} because we choose to number the unique volumes
$w_k$ such that $w_0=\nu_0$.  The labeling of the $w_k$ will therefore agree
with that of the complete path in the case that all volumes in the path are
distinct.  However, in some situations there may be some virtue to ditching
this correspondence and instead choosing to order the $w_i$ from smallest to
largest, or something of that nature, in which case the $w_i$ will be a list
of consecutive volumes.  One must be cautious with this notation, recognizing
that the set $\{w_i\}$ is specific to each individual path.  \emph{Given} the
path $(\nu_{M},\nu_{M-1},\ldots,\nu_{1},\nu_{0})$, one must then
\emph{calculate} $p$ and the corresponding $\{w_i\}$ and their degeneracies
$\{d_i\}$, and then proceeed to evaluate the corresponding path amplitude
(which depends on \emph{all three} sets of numbers.)
} %
The degeneracy of each volume $w_k$ in the given path will be denoted $d_k$,
so $\sum_{k=0}^{p-1}d_k=M+1$. 

The  Hadamard propagator can then be written as 
\begin{equation} \label{eq:vertexexp}
G^\pm_{\mathrm{H}}(\nu_f,\phi_f;\nu_i,\phi_i) =
\sum_{M=0}^{\infty} \sum_{ \substack{\nu_{M-1},\ldots,\nu_1\\ \nu_m\neq \nu_{m+1}}}
A^{\pm}_M(\nu_{M},\nu_{M-1},\ldots,\nu_{1},\nu_{0};\Delta\phi) ,
\end{equation} 
% \begin{eqnarray*}
% G^\pm_{\mathrm{H}}(\nu_f,\phi_f;\nu_i,\phi_i) & = &
% \sum_{M=0}^{\infty} \sum_{{\nu_{M-1},\ldots,\nu_1}\atop{\nu_m\neq \nu_{m+1}}}
% A^{\pm}_M(\nu_{M},\nu_{M-1},\ldots,\nu_{1},\nu_{0};\Delta\phi)   \\
%  & = &
% \sum_{M=0}^{\infty} \sum_{{\nu_{M-1},\ldots,\nu_1}}
% P^{\pm}_M(\nu_{M},\nu_{M-1},\ldots,\nu_{1},\nu_{0};\Delta\phi)
% \end{eqnarray*}
where the ``path amplitude'' $A^{\pm}_M$ associated with the path
$(\nu_{M},\nu_{M-1},\ldots,\nu_{1},\nu_{0})$ is given in terms of the matrix 
elements $\Theta_{\nu\nu'}=\melt{\nu}{\Theta}{\nu}$ by  \cite{sflqc1}%
\footnote{There is a similar alternative expansion for $G^\pm_{\mathrm{NW}}$
in terms of the matrix elements of $H=\sqrt{\Theta}$.
} %
\begin{multline}
A^{\pm}_M(\nu_{M},\nu_{M-1},\ldots,\nu_{1},\nu_{0};\Delta\phi)  =
\Theta_{\nu_M\nu_{M-1}}\ldots\Theta_{\nu_2\nu_1}\Theta_{\nu_1 \nu_0}  \\
\times\prod_{k=0}^{p-1}\frac{1}{(d_k-1)!} \left(\frac{\partial}{\partial \Theta_{w_k w_k}}\right)^{d_k-1}
\sum_{l=0}^{p-1} \frac{e^{\pm i\sqrt{\Theta_{w_l w_l}}\Delta\phi}}{2\sqrt{\Theta_{w_l w_l}}
%        \prod_{ {j=0}\atop{j\neq l}}^{p-1} (\Theta_{w_l w_l}-\Theta_{w_j w_j})}.
       \prod_{ \substack{j=0\\ j\neq l}}^{p-1} (\Theta_{w_l w_l}-\Theta_{w_j w_j})}.
\end{multline}
%
% \begin{eqnarray*}
% G^\pm_{\mathrm{NW}}(\nu_f,\phi_f;\nu_i,\phi_i) & = &
% \sum_{M=0}^{\infty} \sum_{{\nu_{M-1},\ldots,\nu_1}\atop{\nu_m\neq \nu_{m+1}}} \Theta_{\nu_M\nu_{M-1}}\ldots\Theta_{\nu_2\nu_1}\Theta_{\nu_1 \nu_0}\prod_{k=0}^{p-1}\frac{1}{(d_k-1)!}
% \\& &\times \left(\frac{\partial}{\partial \Theta_{w_k w_k}}\right)^{d_k-1} \sum_{m=0}^{p-1} \frac{e^{\pm i\sqrt{\Theta_{w_m w_m}}\Delta\phi}}{\prod_{{j=0}\atop{j\neq m}}^{p-1} (\Theta_{w_m w_m}-\Theta_{w_j w_j})}
% \end{eqnarray*}
%
% (Note the  $\pm$ in the exponent is opposite that of Eq.\ (50) of \cite{cgo11},
% which we believe is a typo.)  

\begin{figure}[hbtp!]
\begin{center}
\subfloat[Spin-foam graph in covariant loop quantum gravity]{
\includegraphics[width=0.55\textwidth]{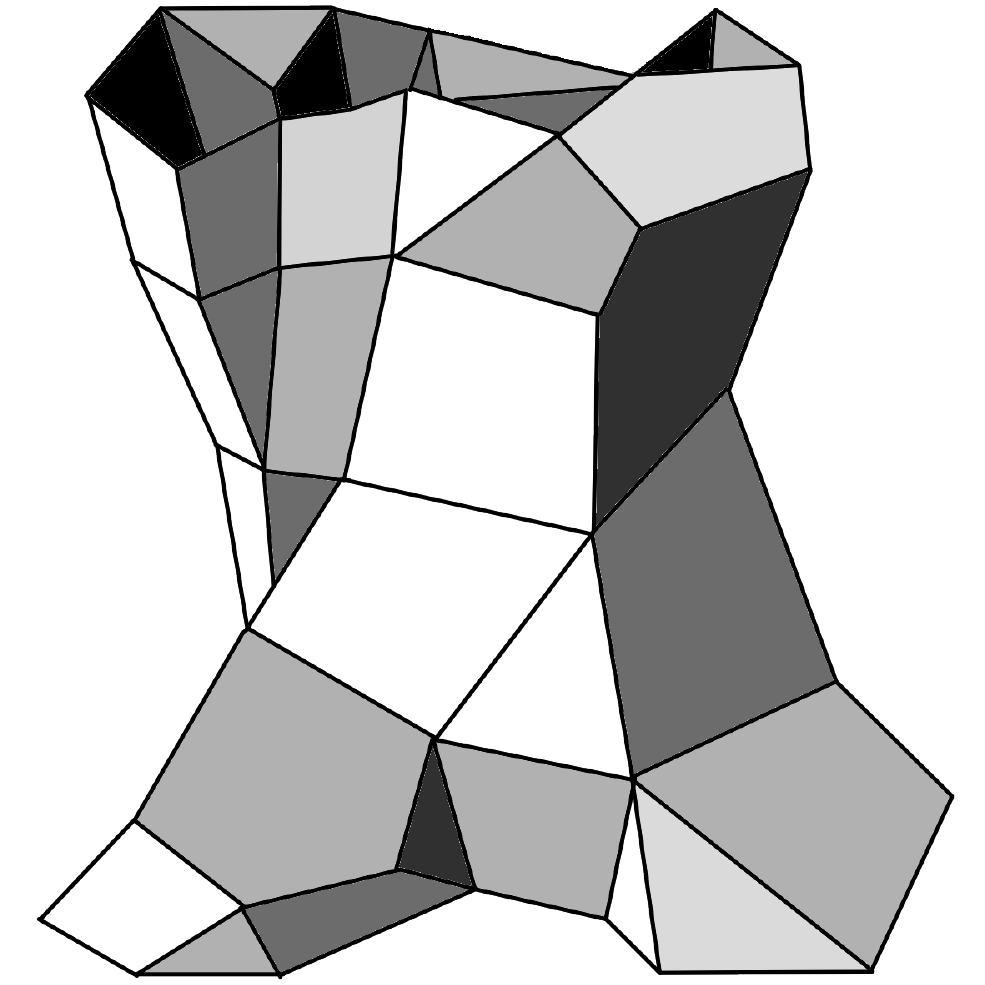}
\label{fig:sfgraph}
}%
% \quad\hspace{1.0cm}
\subfloat[``Spin-foam'' graph on $\mathcal{I}\times\mathcal{V}$ in LQC]{
\includegraphics[width=0.25\textwidth]{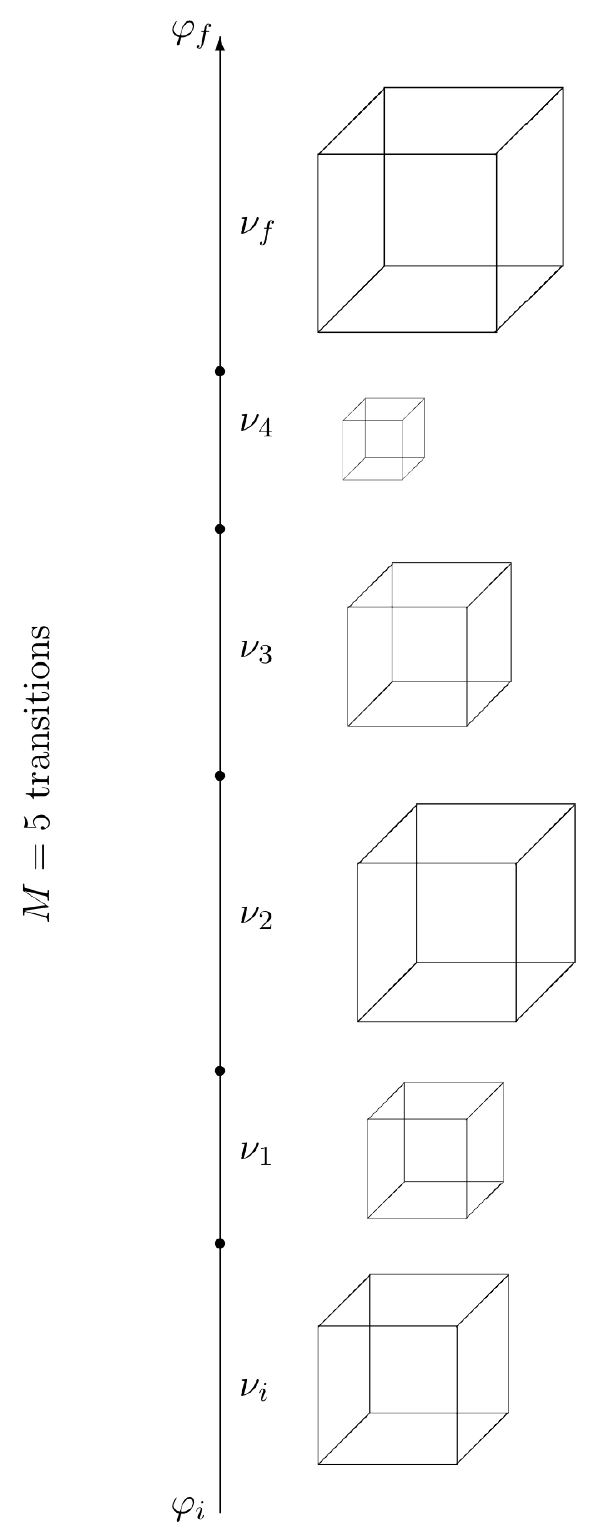}
\label{fig:volhist}%
}%
\end{center}
\vspace{-15pt} %
\caption{In covariant loop quantum gravity, amplitudes for transitions between
fixed ``initial'' and ``final'' boundary surfaces are defined by a
sum-over-histories prescription.  The spacetime manifold is given a (dual)
triangulation with edges colored by spins and an amplitude is assigned to each
such colored dual triangulation.  The full transition amplitude is then
calculated a la Feynman by summing over all possible colorings and (dual)
triangulations -- the ``vertex expansion'' of the amplitudes of spin-foam loop
quantum gravity \cite{RovVid}.  An analogous graph is shown for spin-foam loop
quantum cosmology.  Shown is an example of a particular cosmological history
with $M=5$ volume transitions.  Here $\nu_i=\nu_0$ and $\nu_f=\nu_5=\nu_M$.
The dots (``vertices'') denote the transitions; the value of $M$ is analogous
to the choice of dual triangulation.  The ``edges'' of the graph are labeled by
the volumes, analogous to the spin colorings.  This is analogous to an
individual graph (triangulated spacetime manifold) connecting fixed boundary
surfaces (here fixed $\nu_i$ and $\nu_f$) in spin-foam loop quantum gravity.
The vertex expansion assigns an amplitude to each such history.  The complete
transition amplitude (equivalently, inner product) is the sum of all such
amplitudes.
}%
\label{fig:spinfoam}%
\end{figure}

Eq.~(\ref{eq:vertexexp}) is the ``vertex expansion'' defining the spin-foam
formulation of loop quantum cosmology in analogy with the sum over amplitudes
for transitions between fixed ``initial'' and ``final'' boundary surfaces in
the covariant spin-foam formulation of loop quantum gravity.  In covariant
LQG, the interpolating manifold is given a triangulation with edges colored by
spins, and an amplitude assigned to each such colored triangulation.  The full
transition amplitude is then given in a sum-over-histories prescription by
summing over all possible colorings and (dual) triangulations.  In spin-foam
LQC, the sum over internal volumes is analogous to the sum over colorings, and
the sum over the number of volume transitions analogous to the sum over dual
triangulations.  (See Fig.\ \ref{fig:spinfoam}.)

It is sometimes convenient to regard the path amplitude for any path not
satisfying the condition that each transition must be to a different volume as
simply zero, rather than restricting the sum in Eq.~(\ref{eq:vertexexptheta}) or
(\ref{eq:vertexexp}).  So instead we could write
\begin{equation} \label{eq:vertexexpP}
G^\pm_{\mathrm{H}}(\nu_f,\phi_f;\nu_i,\phi_i) =
\sum_{M=0}^{\infty} \sum_{{\nu_{M-1},\ldots,\nu_1}}
P^{\pm}_M(\nu_{M},\nu_{M-1},\ldots,\nu_{1},\nu_{0};\Delta\phi)
\end{equation} 
as a sum over \emph{all} paths connecting $\nu_i=\nu_0$ to $\nu_f=\nu_M$,
where now
\begin{eqnarray} \label{Pm}
P^{\pm}_M(\nu_{M},\nu_{M-1},\ldots,\nu_{1},\nu_{0};\Delta\phi)  &=& \nonumber 
\Omega_{\nu_M\nu_{M-1}}\ldots\Omega_{\nu_2\nu_1}\Omega_{\nu_1 \nu_0}  \\
&& \hspace{-1cm} 
\times\prod_{k=0}^{p-1}\frac{1}{(d_k-1)!} \left(\frac{\partial}{\partial \Theta_{w_k w_k}}\right)^{d_k-1}
\sum_{l=0}^{p-1} \frac{e^{\pm i\sqrt{\Theta_{w_l 
w_l}}\Delta\phi}}{2\sqrt{\Theta_{w_l w_l}}
\prod_{ \substack{j=0\\ j\neq l}}^{p-1} 
% \prod_{{j=0}\atop{j\neq l}}^{p-1} 
(\Theta_{w_l w_l}-\Theta_{w_j w_j})}.
\end{eqnarray} 
Here we have defined the off-diagonal part of $\Theta$ as $\Omega =
\Theta-\Theta^D$, encompassing the off-diagonal ``transition'' matrix elements
in the vertex expansion, which as we will see gives a non-zero contribution to
Eq.~(\ref{Pm}) only for paths for which all transitions are to distinct,
neighboring volumes.  The diagonal part is denoted by $\Theta^D$.

The case $M=0$ (no transitions) is slightly special.  In this case
$A_0^{\pm}=P_0^{\pm}$, and can be calculated directly to be
\begin{equation} \label{eq:P0}
P_0^\pm (\nu_f,\nu_0;\Delta\phi) = \frac{e^{\pm i \sqrt{\Theta_{\nu_0\nu_0}} 
\Delta\phi}}{2 \sqrt{\Theta_{\nu_0\nu_0}}} 
 \delta_{\nu_f,\nu_0}.
\end{equation} 
(Recall that in our notation $\nu_0=\nu_i$ and $\nu_M=\nu_f$.)

\section{Features of vertex amplitudes in spin foam loop quantum cosmology}
\label{sec:vamp}

\subsection{Vertex amplitudes and their calculation}
\label{sec:vampcalc}

% Section contains: (amplitude)=(path amplitude factor)$\times$(coherence
% factor); expression in terms of $n$ instead of $\nu$; diagrammatic calculation
% method

In order to simplify investigation of various properties of the propagator,
recalling $\nu=4\lambda n$ it is useful to express the matrix elements
$\langle \nu|\Theta| \nu' \rangle = \Theta_{\nu,\nu'}$ of
Eq.~(\ref{eq:thetavv}) in terms of $n$ instead of $\nu$.  That is,
\begin{eqnarray} 
\Theta_{n,n'}
% = 12\pi G \sqrt{|n\cdot n'|}|n+n'|\cdot
&=& \nonumber \kappa^2 \sqrt{|n\cdot n'|}|n+n'|\cdot
 \left\{ \delta_{n,n'}-\frac{1}{2}\left[\delta_{n,n'+1}+\delta_{n,n'-1}  \right]\right\} \\
% &= 2\kappa^2n^2
% \delta_{n,n'}-\kappa^2n^2\left[f_+(n)\delta_{n,n'-1} + f_-(n)\delta_{n,n'+1}\right],
% \quad\text{where}\quad \nu=4\lambda n, \nu'=4\lambda n'
&=& 2\kappa^2n^2
\delta_{n,n'}-\kappa^2n^2\beta(n,n'),
\end{eqnarray} 
where 
\begin{eqnarray}
\beta(n,n') & = &
  \frac{1}{2}\sqrt{\left|\frac{n'}{n}\right|}\left|1+\frac{n'}{n}\right|
      \left[\delta_{n,n'+1}+\delta_{n,n'-1}\right]  \nonumber \\
   & = & f_+(n)\delta_{n,n'-1} + f_-(n)\delta_{n,n'+1},
\end{eqnarray}
with
\begin{equation} 
f_{\pm}(n) =\sqrt{1\pm\frac{1}{n}}\,(1\pm\frac{1}{2n}) ~.
\end{equation} 

It is to be noted that the only non vanishing matrix elements are
$\Theta_{n,n}=2\kappa^2n^2, ~\mathrm{and~} \Theta_{n,n\pm 1} =
-\kappa^2n^2f_{\pm}(n)$.  In the vertex expansion, the off-diagonal matrix
elements corresponding to transitions to distinct volumes can then be written
as
\begin{eqnarray}
\melt{n}{\Omega}{n'}
% = \Omega_{\nu,\nu'} = \Omega_{n,n'}
&=& -\kappa^2 \frac{1}{2}\sqrt{|n\cdot n'|}|n+n'|\cdot
     \left[\delta_{n,n'+1}+\delta_{n,n'-1}  \right] \nonumber \\
&=& -\kappa^2n^2 \beta(n,n') \nonumber\\
&=& -\kappa^2n^2
\left[f_+(n)\delta_{n,n'-1}+f_-(n)\delta_{n,n'+1}\right].
% \quad\text{where}\quad \nu=4\lambda n, \nu'=4\lambda n'
\end{eqnarray}
All matrix elements $\Omega_{n,n'}$ are zero except for $\Omega_{n,n\pm 1} =
-\kappa^2n^2f_{\pm}(n)$.  Crucially, this means that the only paths which have
non-zero amplitude are ones for which each transition is to a neighboring
volume precisely one unit greater or smaller than the one before.  Thus, the
product of off-diagonal terms can be written as
\begin{eqnarray}  \label{eq:Omn} 
% \Omega_{\nu_M\nu_{M-1}}\ldots\Omega_{\nu_2\nu_1}\Omega_{\nu_1 \nu_0} &=& 
% %    (-1)^M \kappa^{2 M} \, \prod_{i=0}^{M-1} n_i^2 \, f_{\pm}(n_i) ~.
%    (-1)^M \kappa^{2 M} \, \prod_{i=0}^{M-1} n_{i+1}^2 \, \beta(n_{i+1},n_i) \nonumber\\
%     &=&  (-1)^M \kappa^{2 M} \, \prod_{i=0}^{M-1} n_{i+1}^2 \, 
%         \prod_{i=0}^{M-1} \left[f_+(n_{i+1})\delta_{n_{i+1},n_i'-1}
%                              +f_-(n_{i+1})\delta_{n_{i+1},n_i'+1}\right] ~.
\Omega_{\nu_M\nu_{M-1}}\ldots\Omega_{\nu_2\nu_1}\Omega_{\nu_1 \nu_0} &=& 
%    (-1)^M \kappa^{2 M} \, \prod_{i=0}^{M-1} n_i^2 \, f_{\pm}(n_i) ~.
   (-1)^M \kappa^{2 M} \, \prod_{i=1}^{M} n_i^2 \, \beta(n_i,n_{i-1}) \nonumber\\
    &=&  (-1)^M \kappa^{2 M} \, \prod_{i=1}^{M} n_i^2 \, 
          \left[f_+(n_i)\delta_{n_i,n_{i-1}-1} + f_-(n_i)\delta_{n_i,n_{i-1}+1}\right] ~.
\end{eqnarray} 
Substituting into Eq.~(\ref{Pm}) and expressing everything in terms of 
$n$ instead of $\nu=4\lambda n$, the path amplitude $P_M^\pm$ becomes
% THIS ONE IS WRONG, I THINK
% \begin{eqnarray} \label{eq:PMn}
% P_M^\pm(n_M,n_{M-1},\ldots,n_1,n_0)
% = \frac{(-1)^M}{2^{2 M + 5/2 - p}} \, \, \prod_{i=0}^{M-1} n_{i+1}^2 \, \beta(n_{i+1},n_i)
% \, \prod_{k=0}^{p-1} \, \frac{1}{(d_k - 1)!} \frac{1}{m_k^{d_k - 1}} \nonumber &&\\ 
%    && \hskip-3.5cm \times  \left(\frac{\partial}{\partial m_k}\right)^{d_k -1} \, 
%        \sum_{l=0}^{p-1} \frac{e^{\pm i \sqrt{2} \kappa m_l \Delta \phi}}{m_l \, 
%         \prod_{{j=0}\atop{j\neq l}}^{p-1} (m_l^2 - m_j^2)} ~,
% \end{eqnarray} 
\begin{eqnarray} \label{eq:PMn}
P_M^\pm(n_M,n_{M-1},\ldots,n_1,n_0) = \nonumber &&\\
  && \hskip-2.5cm  \frac{(-1)^M}{2^{2 M - p + 2}\sqrt{2}\kappa} \, \, 
    \prod_{i=1}^{M} n_i^2 \, \beta(n_i,n_{i-1}) \prod_{k=0}^{p-1} \, \frac{1}{(d_k - 1)!}   
       \left(\frac{1}{m_k}\frac{\partial}{\partial m_k}\right)^{d_k -1} \, 
        \sum_{l=0}^{p-1} \frac{e^{\pm i \sqrt{2} \kappa m_l \Delta \phi}}{m_l \, 
          \prod_{ \substack{j=0\\ j\neq l}}^{p-1} (m_l^2 - m_j^2)} ~,
%           \prod_{{j=0}\atop{j\neq l}}^{p-1} (m_l^2 - m_j^2)} ~,
\end{eqnarray} 
where we have used $\tfrac{\p}{\p \Theta_{nn}} = (4 \kappa^2 n)^{-1}
\tfrac{\p}{\p n}$.  Here we have expressed the full path of $M+1$ volumes as\\
$(\nu_M,\nu_{M-1},\ldots,\nu_1,\nu_0) =
4\lambda\times(n_M,n_{M-1},\ldots,n_1,n_0)$, and the set of $p$ unique volumes
appearing in that path as $(w_{p-1},w_{p-2},\ldots,w_{1},w_{0}) =
4\lambda\times(m_{p-1},m_{p-2},\ldots,m_{1},m_{0})$.  For paths with no
degeneracies (all $d_k=1$, so $p=M+1$) -- the so-called ``Wheeler-DeWitt''
paths -- this simplifies to
\begin{equation} \label{eq:PMnWdW}
P_M^\pm(n_M,n_{M-1},\ldots,n_1,n_0) = 
  \frac{(-1)^M}{2^{M + 1}\sqrt{2}\kappa} \, \, 
    \prod_{i=1}^{M} n_i^2 \, \beta(n_i,n_{i-1})  
        \sum_{l=0}^{M} \frac{e^{\pm i \sqrt{2} \kappa m_l \Delta \phi}}{m_l \, 
          \prod_{ \substack{j=0\\ j\neq l}}^{M} (m_l^2 - m_j^2)} ~,
%           \prod_{{j=0}\atop{j\neq l}}^{M} (m_l^2 - m_j^2)} ~,
\end{equation} 

It is perhaps worth observing that for even modest values of $n$,
$f_{\pm}(n)\approx 1$.  Moreover, since at most one of the two delta functions
in $\beta(n_i,n_{i-1})$ can be non-zero on any section of any path, the
factors of $\beta$ are appreciably different from 1 on supported paths only
for small $n$ (cf.~Eq.~(\ref{eq:Omn})), and therefore except at Planck-scale
volumes essentially serve only to enforce the neighbor-neighbor volume
transitions.

% \vspace{10pt}
% DIAGRAMMATIC CALCULATION METHOD??
% factor of n^2 for each edge/vertex
% path coherence factor like an intertwiner?

\subsection{Structure of volume histories in the vertex expansion}
\label{sec:volhiststruct}

\begin{figure}[hbtp!]
\begin{center}
\includegraphics[width=1.0\textwidth]{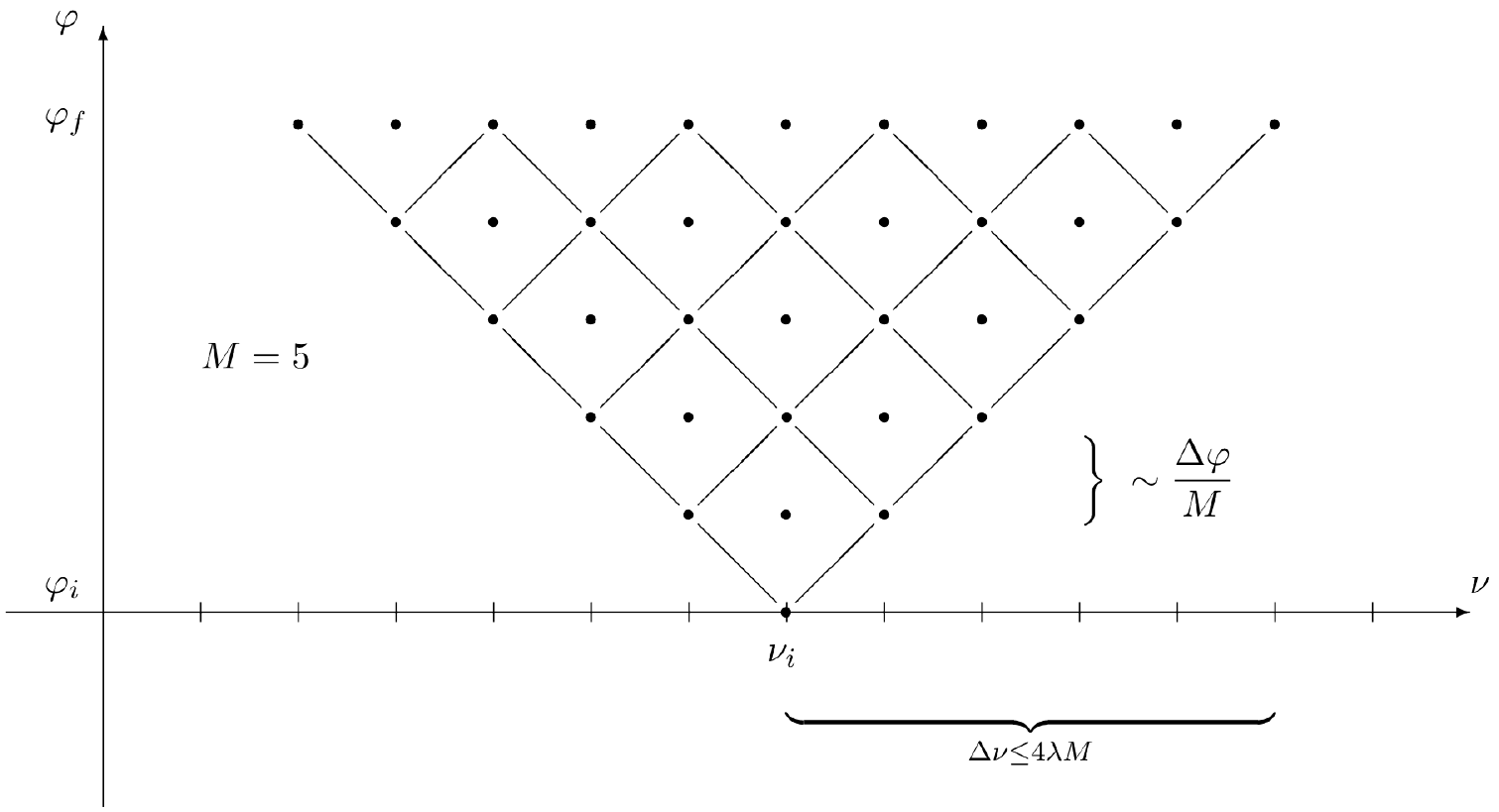}
\end{center}
\vspace{-20pt} %
\caption{The set of all possible cosmological histories that have non-zero
amplitudes starting from volume $\nu_i=4\lambda n_i$ with $M=5$ volume
transitions.  At order $M$ in the vertex expansion, there are $2^M$ possible
distinct supported paths emanating from a given $\nu_i$.  Note that the path
amplitudes assign non-zero values only to paths which connect adjacent
volumes.  Accordingly, the maximum change in cosmological volume at order $M$
in the vertex expansion is bounded above, $\Delta\nu \leq 4\lambda M$ ($\Delta
n \leq M$).  In spite of appearances in the figure, the ``times'' ($\varphi$)
at which the transitions occur are not determined, though the
\emph{``typical''} interval will be of order $\Delta\varphi/M =
(\varphi_f-\varphi_i)/M$.  As we will show in Sec.~\ref{sec:vaqual}, this
implies additionally that the vertex expansion must be carried to orders
$M\gtrsim n\kappa\Delta\varphi$ in order for the expansion to include
trajectories sufficiently fine-grained to accurately capture the relevant
dynamics.
An additional feature of the set of supported paths is that at each ``time''
step (that is, after a given number of transitions), not all possible volumes
are accessible.  The fact that the volume changes by $\pm 4\lambda$ at every
transition means that after an odd/even number of transitions, only volumes
that are an odd/even multiple of $4\lambda$ different from $\nu_i$ are
encountered on a supported path.  
}%
\label{fig:paths}%
\end{figure}

The restriction on the paths arising from the quantum constraint that each
transition is to a neighboring volume only implies that the supported paths
are the $2^M$ ``tree'' graphs starting from $\nu_i=\nu_0$, % 
% \footnote{Strictly speaking, the counting of paths is always this simple only
% if one includes negative $\nu$'s.
% } %
as illustrated in Fig.~\ref{fig:paths}.  Therefore the list of $p$ unique
volumes $(w_{p-1},w_{p-2},\ldots,w_{1},w_{0})$ appearing in a given path will
always be a listing of a consecutive/contiguous range of volumes.  Moreover,
this implies that at order $M$ in the vertex expansion, the absolute
\emph{maximum} change in volume that can be captured at that order is bounded
above, $\Delta\nu \leq 4\lambda M$ ($\Delta n \leq M$).  Among other things,
this immediately implies that very high orders in the vertex expansion will be
necessary to accurately capture cosmological dynamics.  Low orders can only
hope to capture highly quantum behaviors.  In conjunction with qualitative
considerations to be discussed later, this will imply a constraint on just how
large $M$ must be in order to faithfully describe cosmological evolution.

Since all paths in the classical theory either expand or contract, paths in
the vertex expansion which steadily either increase or decrease in volume --
the paths on the boundaries of the tree of possible paths emanating from
$\nu_i$ -- may be referred to as ``Wheeler-DeWitt'' or ``classical'' paths.

As has been emphasized, because of the nearly diagonal structure of the matrix
elements of $\Theta$ (or $\Omega$), only transitions to \emph{neighboring}
volumes have non-zero path amplitudes, enforced by the factors of $\beta$ in
Eq.~(\ref{eq:PMn}).  This is a direct consequence of the structure of the
quantum evolution operator $\Theta$ which only links neighboring volumes in a
uniform discrete grid.
% , or in $n = 4$.  
It is important to recall that this structure followed from restricting to
$j=1/2$ in the trace over $SU(2)$ connections in the field strength tensor in
constructing the quantum the Hamiltonian constraint \cite{aps3}.  If higher
$j$'s are included, we expect the quantum constraint to be increasingly
non-local in volume.%
\footnote{Evidence of this structure exists in the case of $j=1$
\cite{kevinj1} and higher spins \cite{abg17}.
} %
(The expansion of the Newton-Wigner propagator, which involves
$\sqrt{\Theta}$, will also be more non-local, but not sufficiently to
fundamentally alter the general conclusions arrived at here \cite{craig-eigen}.)

\subsection{Volume scaling of vertex amplitudes}
\label{sec:vampscale}

For later use, we would like to understand how the path amplitudes at each
order in the vertex expansion scale with volume ($n$).  To that end, let us
begin by writing them down explicitly for the first few orders in the vertex
expansion (small $M$).

For $M=0$, no degeneracy is possible, $d_0 = 1$. From 
Eq.~(\ref{eq:P0}),
\begin{equation}  \label{eq:P0n}
P_0^\pm (n_0) = \frac{1}{2 \sqrt{2} \kappa}  \frac{e^{\pm i \sqrt{2} \kappa n_o \Delta \phi}}{n_0} 
 \delta_{n_f,n_0}.
\end{equation} 
(Recall that $n_0=n_i$ and $n_M=n_f$.) 

For the case $M=1$, again $d_0 = d_1 = 1$, and
\begin{equation} 
P_1^\pm (n_1,n_0) = -\frac{1}{2^2 \sqrt{2} \kappa} 
n_1^2 \beta(n_1,n_0)
% n_0^2 f_+(n_0) 
\Bigg[\frac{e^{\pm 
\sqrt{2} i \kappa n_0 \Delta \phi}}{n_0(n_0^2 - n_1^2)} + \frac{e^{\pm 
\sqrt{2} i \kappa n_1 \Delta \phi}}{n_1(n_1^2 - n_0^2)} \Bigg]  ~.
\end{equation} 
Since $f_{\pm}(n) \approx 1\pm \tfrac{1}{n}$ for large $n$,
\begin{equation}
n_1^2\beta(n_1,n_0) \approx n_0^2 
  \left[\left(1+\frac{1}{2n_0}\right)\delta_{n_1,n_0-1}+\left(1-\frac{1}{2n_0}\right)\delta_{n_1,n_0+1}  \right]  ~.
\label{eq:n2beta}
\end{equation}
Therefore, on supported paths (so $n_1=n_0\pm1$), $P_1^{\pm}$ scales at large
$n_0$ as
\begin{equation} 
P_1^\pm (n_1,n_0) \approx -\frac{\pm}{2^3 \sqrt{2} \kappa} \left(1 \mp \frac{1}{2n_0}\right)  
  e^{\pm \sqrt{2} i \kappa n_0 \Delta \phi} 
     \left[1-  \frac{e^{\mp \sqrt{2} i \kappa \Delta \phi}}{1\mp\frac{1}{n_0}} \right]  ~.
\end{equation} 
(The signs depend on the branch of the supported paths you're on.) 
For $M=2$, all $d_k = 1$ (the ``Wheeler-DeWitt'' paths), one finds 
\begin{multline} 
P_2^\pm (n_2,n_1,n_0) = \frac{1}{2^3 \sqrt{2} \kappa}  \, 
n_2^2 \beta(n_2,n_1)\, n_1^2\beta(n_1,n_0) ~\times   \\
% n_0^2 n_1^2 f_\pm(n_0) f_\pm(n_1) 
\Bigg[\frac{e^{\pm i \sqrt{2} \kappa n_0 \Delta \phi}}{n_0(n_0^2 - n_1^2) (n_0^2 - n_2^2)} 
+ \frac{e^{\pm i \sqrt{2} \kappa n_1 \Delta \phi}}{n_1(n_1^2 - n_0^2) (n_1^2 - n_2^2)} 
      + \frac{e^{\pm i \sqrt{2} \kappa n_2 \Delta \phi}}{n_2(n_2^2 - n_0^2) (n_2^2 - n_1^2)} \Bigg] ~.
\end{multline} 
When $n_0$ is large, on supported Wheeler-DeWitt paths ($n_1=n_0\pm1$, 
$n_2=n_0\pm2$), we obtain similarly
\begin{eqnarray} 
P_2^\pm (n_2,&&n_1,n_0)\approx \frac{1}{2^3 \sqrt{2} \kappa} n_0 \left(1 \pm \frac{1}{1 \pm n_0}\right) \left(1 \pm \frac{1}{n_0}\right)^3  e^{\pm i \sqrt{2} \kappa n_0 \Delta \phi} \times \nonumber \\
&& \Bigg[\frac{1}{\left(1 \pm \frac{2}{n_0}\right) \left(1 \pm 
\frac{4}{n_0}\right)} - \frac{e^{\pm i \sqrt{2} \kappa \Delta \phi}}{\left(1 
\pm \frac{1}{n_0}\right) \left(\pm 1 + \frac{1}{n_0}\right) \left(\pm 2 + 
\frac{1}{n_0}\right)} +  \frac{e^{\pm i 2 \sqrt{2} \kappa \Delta \phi}}{2 \left(1 \pm \frac{2}{n_0}\right) \left(\pm 1 + \frac{1}{n_0}\right) \left(\pm 4 + \frac{1}{n_0}\right)}\Bigg]  ~.
\end{eqnarray} 
(The various $\pm$ signs in this relation are mostly uncorrelated with one
another.)
 
We see that $P_M^{\pm}$ for the Wheeler-DeWitt paths with $M=0, 1, 2$ scale
with volume like $n_0^{M-1}$ in the large-$n_0$ limit.  This scaling with
volume may at first seem surprising, because one might conclude by naive
counting of powers of $n$ and $m$ in Eq.~(\ref{eq:PMn}) that the $P_M^{\pm}$
scale roughly as $1/n$ with volume.  However, the situation is a bit more
subtle than that.  Without becoming bogged down in the details, the subtleties
arise because of the difference of squares in the denominator of
Eq.~(\ref{eq:PMn}) and the fact that the unique volumes appearing in the list
for each supported path are consecutive.  Therefore that difference always
contains terms which are close neighbors, which therefore scale as $n$ rather
than $n^2$, as taken into account explicitly in the calculations above.  This
alters the conclusions drawn from power-counting.
 
Indeed, the pattern of volume scaling $n_0^{M-1}$ of the $P_M^{\pm}$ for the
Wheeler-DeWitt paths in the large-$n_0$ limit continues for all $M$.  To see
this, we need to assess the volume scaling of both volume-dependent factors in
Eq.~(\ref{eq:PMnWdW}).  Let us begin with the simpler of the two, the product
of transition matrix elements that leads to $\prod n^2\beta$.  The
expanding/contracting Wheeler-DeWitt paths can be indexed as $n_i=n_0\pm i$,
$i=0\ldots M$.  Then
\begin{eqnarray}
\prod_{i=1}^{M} n_i^2 \beta(n_i,n_{i-1}) & = & 
  \prod_{i=1}^{M} (n_0+i)^2
  \prod_{j=1}^{M} f_{\mp}(n_j)
\nonumber\\
 & = &   n_0^{2M} \,
  \prod_{i=1}^{M} \left(1\pm\frac{i}{n_0}\right)^2
  \prod_{j=1}^{M} f_{\mp}(n_0(1\pm j/n_0)).
\label{eq:pathamp}
\end{eqnarray}
Now,
\begin{equation}
 \prod_{i=1}^{M} \left(1\pm\frac{i}{n_0}\right)^2 = 
(\pm)^M  \frac{1}{n_0^M}\frac{\Gamma(M+1\pm n_0)}{\Gamma(1\pm n_0)}.
\end{equation}
Using Stirling's formula%
\footnote{As well as $\lim_{n\rightarrow\infty}(1+x/n)^n = e^x$, and the
reflection identity for the case of contracting paths.
} %
it is straightforward to show that this approaches unity in the limit $n_0 \gg
M$.  In a similar manner, the product of $f_{\mp}$ (recalling
$\prod_i\sqrt{x_i}=\sqrt{\prod_i x_i}$) also approaches unity in the same
limit, and as expected Eq.~(\ref{eq:pathamp}) scales as $n_0^{2M}$ in the
large-$n$ limit.

The volume scaling of the sum appearing in Eq.~(\ref{eq:PMnWdW}) is determined
by the volume products in the denominators.  For simplicity we will imagine an
expanding Wheeler-DeWitt path, but the choice doesn't actually make any
difference to the analysis.  Let us concentrate on a single such factor for an
arbitrary choice of $l$, where of course $0 \leq l \leq M$.  We parameterize
the volumes appearing in the product relative to $n_l=n_0+l$ by a set of
integers $k_j^l$, so that $n_j=n_l+k_j^l$, where $k_j^l\in \{
-l,-l+1,\ldots,-1,1,\ldots, M-l \}$.  In that case the volume products in the
denominators are
\begin{eqnarray}
n_l\, \prod_{ \substack{j=0\\ j\neq l} }^{M} (n_l^2 - n_j^2) & = & 
  n_l\, \prod_{j=0}^{l-1} (n_l^2 - n_j^2) \prod_{j=l+1}^{M} (n_l^2 - n_j^2)    
\nonumber\\
 & = &  n_l^{M+1} (-2)^M 
  \prod_{j=0}^{l-1} k_j^l  \prod_{j=l+1}^{M}  k_j^l
   \prod_{j=0}^{l-1}  \left(1+\frac{k_j^l}{2n_l}\right)   
   \prod_{j=l+1}^{M} \left(1+\frac{k_j^l}{2n_l}\right)
\nonumber\\
 & = &  n_l^{M+1} (-2)^M 
   (-)^l \prod_{j=1}^{l} j  
   \prod_{j=1}^{M-l}  j
   \prod_{j=1}^{l}  \left(1-\frac{j}{2n_l}\right)   
   \prod_{j=1}^{M-l} \left(1+\frac{j}{2n_l}\right)
\nonumber\\
 & = &  n_l^{M+1} (-2)^M 
(-)^l\Gamma(l+1)  \Gamma(M-l+1)
\frac{1}{(2n_l)^l} \frac{\Gamma(2n_l)}{\Gamma(2n_l-l)}
\frac{1}{(2n_l)^{M-l}} \frac{\Gamma(M-l+1+2n_l)}{\Gamma(1+2n_l)}
\nonumber\\
 & = &  (-)^{M+l}
\Gamma(l+1)  \Gamma(M-l+1) \, n_l \, 
\frac{\Gamma(2n_l)}{\Gamma(2n_l-l)}
\frac{\Gamma(M-l+1+2n_l)}{\Gamma(1+2n_l)}  ~.
\label{eq:voldenom}
\end{eqnarray}
The crucial point is now that application once again of Stirling's formula to
all of the factors involving $n_l$ in this expression reveals that the volume
denominators Eq.~(\ref{eq:voldenom}) scale as $n_l^{M+1} \sim n_0^{M+1}$ in
the limit $n_l\gg M$.  Thus, as was to be shown, the \emph{ratios} of the
leading transition prefactor Eq.~(\ref{eq:pathamp}) to each of the volume
denominators Eq.~(\ref{eq:voldenom}) scale as $n^{2M}/n^{M+1} = n^{M-1}$ in
the limit $n\gg M$, rather than the $1/n$ that would be expected from naive
power-counting.%
\footnote{The underlying reason for the difference is now actually easy to
see.  In each of the $M$ difference-of-squares factors in the product,
$n_l^2-n_j^2 = (n_l+n_j)(n_l-n_j)$.  In the limit $n\gg M$, all of the $M$
difference terms scale like $n$ rather than $n^2$, and so the power counting
should properly give $1/n \times n^M = n^{M-1}$ in that limit.  The detailed
analysis bears out this expectation.
} %

These considerations can become quite involved, and we do not attempt to 
repeat this analysis for the (much more numerous) paths with degeneracies 
here.  The arguments we make for the Wheeler-DeWitt paths are sufficient for 
our purposes below.

\subsection{Satisfaction of the constraint in the vertex expansion}
\label{sec:constraint}
An important property of the Hadamard propagator is that it satisfies the
constraint, or in other words, the action of $\hat C$ on the propagator
vanishes:
\begin{equation}  \label{CGeq}
\hat C G^\pm_H(\nu_f, \phi_f; \nu_i, \phi_i) = 0 . 
\end{equation} 
This is easily seen from, for example, Eq.~(\ref{eq:GH}).  Can it also be
shown from the vertex expansion?  The answer is that it can.  However, we
shall argue that satisfaction of the constraint strictly holds \emph{only} if
one includes \emph{all} the terms in the vertex expansion from $M =
0...\infty$.  A pertinent question is to ask in what sense a \emph{truncated}
series in $M$ is a solution of the quantum Hamiltonian constraint.  In
Ref.~\cite{sflqc2}, this issue was addressed using a bookkeeping perturbation
parameter $(\lambda)$ introduced via $\Theta = \Theta^D + \lambda \Omega$.
(Their bookkeeping $\lambda$ is distinct from the $\lambda$ related to the area
gap we use in this paper.)  It was then shown that (\ref{CGeq}) is satisfied
% order by order 
in the vertex expansion \cite{sflqc2} in the sense that
\begin{equation} \label{m-eq}
\left(\partial_\phi^2 +\Theta^D\right)P_M(n_f,\phi_f;n_i,\phi_i)
     + \Omega P_{M-1}(n_f,\phi_f;n_i,\phi_i) = 0     ~,
\end{equation} 
so that the diagonal piece of the constraint at any order $M$ in the vertex
expansion is cancelled by the off-diagonal piece from one order down.  Here
the $P_M$ appearing in these expressions and following are the \emph{sums}
over all paths at order $M$ of the individual path amplitudes appearing in
Eq.~(\ref{eq:vertexexpP}).  That is,
\begin{equation}   \label{eq:PMsum}
P_M(\nu_f,\phi_f;\nu_i,\phi_i) = 
\sum_{{\nu_{M-1},\ldots,\nu_1}} 
 P^{\pm}_M(\nu_{M}=\nu_f,\nu_{M-1},\ldots,\nu_{1},\nu_{0}=\nu_i;\Delta\phi) ~.
\end{equation}
If the maximum number of volume transitions is $M^*$, then in \cite{sflqc2} it
is argued that
\begin{equation} \label{m-eq-trunc}
\left(\partial_\phi^2 +\Theta^D + \Omega \right) \sum_{M=0}^{M^*} \lambda^M P_M(n_f,\phi_f;n_i,\phi_i) 
       + \Omega P_{M-1}(n_f,\phi_f;n_i,\phi_i) = {\cal O}{(\alpha^{M^* + 1})} ~.
\end{equation} 
Here the term which is approximated away is the off-diagonal piece originating
from $\Omega$ acting on the path amplitude for the $M^*$th volume transition.
Though this relation appears to suggest that the constraint is satisfied to
order $M^*$, it is important to note that $\lambda$ is not small.  In fact, as
a bookkeeping parameter it is strictly equal to unity \cite{sflqc2}.  It is
therefore unclear what is the error in truncating the vertex expansion at
finite $M^*$.

To understand the truncated vertex expansion and whether it satisfies the
constraint within a controlled approximation, let us begin with the $M=0$
term.  In this case, the off-diagonal term $\Omega$ is trivially zero and
Eq.~(\ref{m-eq}) yields
\begin{equation} \label{m0eq}
\left(\p_\phi^2 + \Theta^D \right) P_0^+(n_1,n_0) = 0 ~.
\end{equation} 
Using the explicit expression of $P_0^+$, it is straightforward to verify that
the L.H.S of (\ref{m0eq}) indeed vanishes:
\begin{equation} 
\p^2_\phi P_0^+(n_1,n_0) = 
 - \frac{1}{\sqrt{2}} \kappa n_0   e^{i \sqrt{2} \kappa n_0 \Delta \phi} \, \delta_{n_1,n_0} 
  = - \sum_{n_1'} \Theta^D_{n_1,n_1'} P_0^+(n_1',n_0) = - \Theta^D 
P_0^+(n_1,n_0)  ~.
\end{equation} 
Hence, the L.H.S of Eq.~(\ref{m0eq}) is identically zero.  The constraint in
this particular case is satisfied exactly since $\Omega = 0$.

Now let us consider Eq.~(\ref{m-eq}) for $M=1$. Then, 
\begin{equation} 
\left(\p_\phi^2 + \Theta^{\tiny{D}} \right) P_1^+(n,n_0) = -\Omega P_0^+ (n,n_0)
\end{equation} 
After a straightforward computation, the L.H.S yields
\begin{eqnarray} \label{m1diageq}
\left(\p_\phi^2 + \Theta^D \right) P_1^+(n_1,n_0) &=& \nonumber \left(\p_\phi^2 + 2 \kappa^2 n_1^2\right) P_1^+(n_1,n_0) \\
&=&  \frac{\kappa}{2\sqrt{2}} \left(\frac{(n_0-1)^2}{n_0} f_+(n_0-1) 
    \delta_{n_1,n_0-1} + \frac{(n_0+1)^2}{n_0} f_-(n_0+1) \delta_{n_1,n_0+1} \right) 
        e^{i \sqrt{2} \kappa n_0 \Delta \phi} .
\end{eqnarray} 
% &=& \nonumber -\frac{\kappa}{2\sqrt{2}} \frac{n_1^2}{n_0} \beta(n_1,n_0) e^{i \sqrt{2} \kappa n_0 \Delta \phi} \\
% &=& -\nonumber \sum_{n_1'} \Omega_{n_1,n_1'} P_0^+(n_1',n_0) \\
% &=& -\nonumber \Omega P_0^+(\nu_1,\nu_0)
% \eq
Similarly, we can compute the off-diagonal part for $M=0$ which contributes in
Eq.~(\ref{m-eq}) for $M^*=1$.  It turns out to be
\begin{eqnarray}  \label{m0offdiageq}
\Omega P_0^+(n_1,n_0) &=& \nonumber \sum_{n_1'} \Omega_{n_1,n_1'} P_0^+(n_1',n_0) \\
&=& \nonumber -\frac{\kappa}{2\sqrt{2}} \frac{n_1^2}{n_0} \beta(n_1,n_0) 
              e^{i \sqrt{2} \kappa n_0 \Delta \phi}\\
&=&-\frac{\kappa}{2\sqrt{2}} \left(\frac{(n_0-1)^2}{n_0} f_+(n_0-1) \delta_{n_1,n_0-1} 
   + \frac{(n_0+1)^2}{n_0} f_-(n_0+1) \delta_{n_1,n_0+1} \right) e^{i \sqrt{2} \kappa n_0 \Delta \phi} ~.
\end{eqnarray} 
Hence Eq.~(\ref{m-eq}) is again satisfied, up to the error term which is the
off-diagonal piece for $M=1$.  To estimate the error involved, let us compute
this term:
\begin{eqnarray} \label{m1offdiageq}
\Omega P_1^+(n_1,n_0) &=& \nonumber -\kappa^2 n_1^2 \sum_{n_1'} \beta(n_1,n_1') P_1^+(n_1',n_0) \\
&=&\nonumber \frac{1}{4 \sqrt{2}} \kappa n_1^2 \bigg[\frac{(n_0-1)^2}{2n_0-1} f_+(n_0-1) \left(\frac{1}{n_0} - \frac{e^{-i  \sqrt{2} \kappa n_0 \Delta \phi}}{n_0-1} \right) \beta(n_1,n_0-1) \\
&& ~~~~~~~~~~~~~~~~~~~~ - \frac{(n_0+1)^2}{2n_0+1} f_-(n_0-1) 
\left(\frac{1}{n_0} - \frac{e^{i  \sqrt{2} \kappa n_0 \Delta \phi}}{n_0+1} \right) \beta(n_1,n_0+1)\bigg] ~.
\end{eqnarray} 
For large $n$ (volume), this scales as $n^2$.  On the other hand, the rest of
the terms (\ref{m1diageq}) and (\ref{m0offdiageq}) scale as $n$.  In
particular, the ratio of the remainder term and any of the terms in
Eq.~(\ref{CGeq}) for $M^*=1$ scales as follows:
\begin{equation} 
\frac{\Omega P_1^+(n_1,n_0) }{(\p_\phi^2 + \Theta^D)P_1^+(n_1,n_0)} \sim n  ~.
\end{equation} 
If truncation at $M^*=1$ was to be viable in general this ratio would need to
be much smaller than unity.  However, the relative error in the truncation
seems to grow with volume $n$.

This is not a phenomenon restricted to low-$M$.  We again restrict attention
to the Wheeler-DeWitt paths to make our point.  Let us examine the scaling
with $n$ for a general truncation order $M$.  We have seen explicitly that for
small $M$, $(\p_\phi^2 + \Theta^D)P_{M}$ scales as $n^M$.  However, $\Omega
P_{M}$ scales as $n^{M+1}$.  Hence their ratio scales as $n$, as was found for
$M = 1$ in the above computation.  This scaling continues at all orders $M$ in
the limit $n\gg M$.  This is, in fact, required by Eq.~(\ref{m-eq}).  We
already know that when $n\gg M$, the $P_M$ scale as $n^{M-1}$.  Then $\Omega
P_{M-1}$ will scale as $n^{M+1}$, and Eq.~(\ref{m-eq}) therefore requires that
$(\p_\phi^2 + \Theta^D)P_{M}$ does as well.  (In other words, Eq.~(\ref{m-eq})
demands that $(\p_\phi^2 + \Theta^D)$ changes the volume scaling of $P_M$ by
one order lower than does $\Omega$.)  Thus, for an arbitrary truncation order
$M^*$, the relative truncation error is
\begin{equation} 
\frac{\Omega P_{M^*}^+ (n_M,n_0) }{(\p_\phi^2 + \Theta^D)P_{M^*}^+ (n_M,n_0)} 
\sim n   ~,
\end{equation} 
at least in the limit $n\gg M$ for Wheeler-DeWitt paths.

This relation implies that there is a regime ($n\gg M$) in which the term
which is ignored can be at least of the same order as the terms in
Eq.~(\ref{m-eq}), at least for some paths in the expansion.  On inclusion of
this term, the constraint does not annihilate the truncated Hadamard
propagator at any given order $M$ in the vertex expansion.%
\footnote{Repeating this exercise in the non-relativistic representation, the
same conclusion is expected to hold for the Newton-Wigner propagator.
} % 
Hence the truncated vertex expansion does not faithfully captures the
vanishing of the quantum constraint: a naively truncated expansion does
\emph{not} appear to be a solution to the constraint at any finite order within a
controlled approximation. %  
% \footnote{There remains a possibility, which we have so far not absolutely
% ruled out, that cancellations in the sum in Eq.~(\ref{eq:PMsum}) somehow
% obviate our arguments.  Note, however, that this does not occur at the low
% values of $M$ we have studied explcitly.
% % Could this really be?   That would mean that degenerate paths would 
% % contribute something that exactly cancels the Wheeler-DeWitt growth, plus 
% % pieces that something small.
% } %
A consequence of this is that the quantum bounce will not necessarily manifest
at any truncated order in the expansion.  We have already seen that large
orders in the vertex expansion are in any event required to capture large
changes in volume.  Hence, for small $M^*$ the quantum bounce may not be
evident in the vertex expansion.

It remains to consider whether it is possible to reorganize the vertex
expansion in spin foam LQC in such a manner as to arrive at an expansion that
may be truncated at finite order and remain a solution to the quantum
constraint in a controlled way.  We do not take up this question here.

\subsection{Qualitative considerations}
\label{sec:vaqual}

% Section contains: good at most for small $\Delta\phi$ at low $M$; probing
% bounce requries large $M$; degenerate paths dominate;

It is important to understand in this ``covariant'' description of LQC to what
extent the vertex expansion discussed in the previous sections can be used as
a tool to probe quantum cosmological dynamics.  For example, can we employ the
vertex expansion to probe the quantum bounce of a loop quantized universe
starting from a macroscopic volume?  Unfortunately, what we find is that in
order to probe long evolutions it is necessary to carry the vertex expansion
to very (very) high orders.

We have already seen that large orders in the vertex expansion are required in
order to capture large changes in volume.  Just how large must the order of 
the expansion be?

Our starting point is the effective Hamiltonian in LQC \cite{aps3}, which for
the case of a spatially flat homogeneous and isotropic spacetime with 
massless scalar matter is given by
\begin{equation} 
C_{\rm{eff}} = p_\phi^2 - \frac{\sqrt{3}}{{4}} \gamma \hbar \nu^2 \sin^2(\lambda b) ~.
\end{equation} 
Solving Hamilton's equations for $\nu$ and $\phi$, we obtain  
\begin{equation} 
\nu = \nu_0 \, \cosh(\kappa(\phi - \phi_0)) ~,
\end{equation} 
where $\nu_0$ and $\phi_0$ are constants of integration. Consequently,
\begin{equation} 
\kappa \d \phi = \frac{\d \nu}{\sqrt{\nu^2 - \nu_0^2} } ~.
\end{equation} 
With $\nu = 4 n \lambda$ and in the limit of large volume $n \gg n_0$, we find
%For $\nu \gg \nu_0$,
\begin{equation} 
\kappa \d \phi \approx n^{-1} \d n ~.
\end{equation} 
This tells us that semiclassical cosmological trajectories require a ``time'' 
interval $d\phi$ of order $1/n \kappa$ in order for the volume to change 
by one Planck unit, $dn=1$.

The vertex expansion expresses amplitudes for transitions between
$(\nu_i,\phi_i)$ and $(\nu_f,\phi_f)$ as a sum of amplitudes for paths with
$M$ Planck-scale volume transitions.  Since these amplitudes are zero unless
the volumes are adjacent, we have already seen that at a minimum, it must be
that $M\geq |\nu_f-\nu_i|/4\lambda = |n_f-n_i|$.%  
\footnote{In point of fact, it must be that $M$ is greater than the largest
difference in volumes exhibited by the trajectory of interest.  For example,
to capture a quantum bounce, $M$ must be greater than ($1/4\lambda$ times) the
difference between the initial volume and the bounce volume.  If the initial
volume is macroscopic, this will be a large $M$ indeed.
} %
Additionally, with $\phi_f-\phi_i = \Delta \phi$, for paths with $M$
transitions the ``typical'' time $\phi$ available for each transition is of
order $\Delta \phi/M$.  In order for the vertex expansion at order $M$ to
effectively capture the dynamics, it must be that the paths included are
sufficiently fine-grained, so that the time required for each transition is
less than $1/n \kappa$.  In other words, it must be that $\Delta \phi/M
\lesssim 1/n \kappa$, so that
\begin{equation} 
M \gtrsim n \kappa \Delta \phi ~.
\end{equation} 
This inequality implies that in order to probe long intervals of cosmological
``time'' $\phi$ using the vertex expansion, we must continue the expansion to
sufficiently large values of $M$.  As a corollary, the vertex expansion at low
orders in $M$ is at best a short-time expansion useful for investigating
highly quantum phenomena.  Therefore, the vertex expansion at low $M$ cannot
be used to accurately probe quantum cosmological dynamics unless
\textbf{\emph{both}} $\Delta\phi$ \emph{and} the change in volume
$|\nu_f-\nu_i|$ are small.

\section{Discussion}
\label{sec:disc}

We have explored the spin-foam-like ``vertex expansion'' of the transition
amplitudes of solvable loop quantum cosmology in some detail, an expansion
that is in many ways analogous to the vertex expansion of covariant spin-foam
loop quantum gravity.  The hope expressed in Refs.~\cite{sflqc1,*sflqc2} is
that insights from an exactly solvable model in the canonical picture might
help shed some light on some of the difficult conceptual issues arising in the
full covariant theory.  Here, we have instead focused on the features and 
utility of the vertex expansion as a tool for investigating LQC itself.

We have found that the vertex expansion has many of the features of an
expansion in $Et/\hbar$ of a standard quantum mechanical propagator.  That is,
that low orders in the expansion can accurately capture the quantum dynamics
only for short ``times'' $\Delta\phi$.  Very large orders in the expansion are
required in order to probe dynamics on cosmologically relevant scales.
Moreover, truncating the expansion at any finite order does not appear to be a
well-controlled approximation, and is likely to obscure evidence of the
marquee feature of LQC, the quantum bounce at small volume, even though the
bounce may be seen clearly in the propagator itself, for example, from
Eq.~(\ref{eq:GH}) \cite{consistent-lqc,CS17c}.  Ironically,
therefore, while the vertex expansion for LQC may be illuminating for certain
conceptual issues, in its current form it is unlikely to prove a useful tool
for direct calculations of cosmologically relevant questions.%  
\footnote{It is worth reiterating that the spin-foam-like vertex expansion we
have been discussing is conceptually distinct from the approach to spin-foam
cosmology from the full covariant theory \cite{sf-cosmo1, sf-cosmo2,
sf-sloan}.
} %
The vertex expansion may, nevertheless, prove useful for investigating highly
quantum, short-time questions of Planck-scale dynamics.

This feature of the vertex expansion has its origins in the essentially local
dynamics, Eq.~(\ref{eq:theta}), of solvable loop quantum cosmology (sLQC).  In
turn, as noted above this highly local dynamics is connected with the
restriction to $j=1/2$ in the quantization of sLQC. The inclusion of higher
spins should yield a dynamics that links more distant volumes.  An example is
the case of the quantum Hamiltonian constraint obtained using the $j=1$
representation in LQC \cite{kevinj1}.  In this particular case, the quantum
evolution operator $\Theta$ links wavefunctions at nine steps in volume with a
maximum non-local volume difference equaling $8 \lambda$ instead of $4
\lambda$.  It is expected that for higher $j$, a more non-local quantum
difference equation will be obtained \cite{abg17}.  In such models, lower
orders $M$ in the corresponding vertex expansion may accurately capture the
dynamics of longer ``time'' intervals.

Alternately, it may turn out to be possible to re-order the vertex expansion
in such a way that low orders in the expansion accurately capture
semi-classical dynamics and key global quantum features such as the bounce --
one that can exhibit large volume changes at low orders in the expansion, and
can be truncated consistently at finite order in a controlled way.  This would
be more akin to the WKB ($\hbar$) expansion in ordinary quantum theory.
Indeed, in keeping with this line of thought, in Ref.~\cite{ach10b} the
transition amplitudes are expressed as an ordinary path integral over phase
space variables, and the semi-classical dynamics and quantum bounce become
evident at lowest order in the standard way (stationary action).  Nonetheless,
a controlled order-by-order expansion could be of some use.  Indeed, it is
likely there exist physically distinct expansions depending on the choice of
order parameter, for example, $l_p$ vs.~$\lambda$, with different physical
meanings and uses, but we do not take up that possibility here.%
\footnote{The expansion in terms of the matrix elements of $\sqrt{\Theta}$
discussed in e.g.~Ref.~\cite{sflqc2}, on the other hand, will have very
similar properties to those discussed in this paper , but will be more
non-local because of the square-root.  }

% \appendix*
% \appendix* % For a single un-named appendix
% \appendix  % For more than one appendix

% \section{Appendix}
% \label{app:}
% \renewcommand{\theequation}{A.\arabic{equation}}

% \section{}
% \label{sec:}

% This appendix contains terribly important technical results.

%\begin{figure}[tbh!]
%\centering
%\includegraphics[scale=0.55]{fig3.eps}
%\caption{Behavior of volume of the universe versus time is shown for a radiation-dominated ($w=1/3$) universe. Initial conditions for %volume and energy density, and designations for these solutions are the same as in Fig. 2.}
%\end{figure}

%\begin{figure}[tbh!]
%\centering
%\includegraphics[scale=0.55]{fig4.eps}
%\caption{$V$ vs. $t$ for a matter-dominated ($w=0$) universe is demonstrated. The choice of $V_0$ and $\rho_0$, and labeling of solutions %is same as in Fig 2.}
%designations for these solutions are the same as in Figure 2.}
%\end{figure}

\section*{Acknowledgments}
D.C.\ would like to thank the Department of Physics and Astronomy at Louisiana
State University, where portions of this work were completed, for its
hospitality.  D.C.\ was supported in part by a grant from FQXi.  PS is
supported by NSF grants PHY-1404240 and PHY-1454832.

% % PARAM'S
% 
% \bibliographystyle{h-physrev5}
% 
% %\bibliography{psbib}%\bibliography{VertexSFLQCNotes}
% 
% %\bibliographystyle{h-physrev5}
% 
% % \bibliography{psbib}
% 
% \bibliography{VertexSFLQC}

% Homegrown bibliography:
% \input{dLQC.bibl}

% Using BibTeX:
% For arXiv or PRD, input only .bbl file; otherwise, process bib:
\bibliographystyle{h-physrev5}
\ifthenelse{\arxiv=1}{%
\bibliography{VertexSFLQC}
}{%
\bibliography{global_macros,../Bibliographies/master,psbib}%
% \bibliography{../Bibliographies/global_macros,../Bibliographies/master}
}% 

\end{document}